\documentclass[3p]{elsarticle}

\usepackage{framed,multirow}

\usepackage{amssymb}
\usepackage{latexsym}
\usepackage{graphicx}

\usepackage{url}
\usepackage{acro}
\usepackage{siunitx}
\usepackage{amsmath}
\usepackage{xcolor}
\definecolor{newcolor}{rgb}{.8,.349,.1}

\usepackage[citebordercolor=white]{hyperref}
\DeclareAcronym{LiAISON}{short = LiAISON, long = Linked Autonomous Interplanetary Satellite Orbit Navigation}
\DeclareAcronym{BER}{short = BER, long = Bit Error Rate}
\DeclareAcronym{CRTBP}{short = CRTBP, long = Circular Restricted Three-body Problem}
\DeclareAcronym{LU}{short = LU, long = Length Unit}
\DeclareAcronym{TU}{short = TU, long = Time Unit}
\DeclareAcronym{EKF}{short = EKF, long =  Extended Kalman Filter}
\DeclareAcronym{STM}{short = STM, long = State Transition Matrix}
\DeclareAcronym{RMS}{short = RMS, long = Root Mean Square}
\DeclareAcronym{LOS}{short = LOS, long = Line-of-Sight}
\DeclareAcronym{OM}{short = OM, long = Observability Matrix}
\DeclareAcronym{SVD}{short = SVD, long = Singular Value Decomposition}
\DeclareAcronym{SNR}{short = SNR, long = Signal-to-Noise}
\DeclareAcronym{FIM}{short = FIM, long = Fisher Information Matrix}
\DeclareAcronym{CRLB}{short = CRLB, long = Cramér–Rao Lower Bound}
\DeclareAcronym{SKF}{short = SKF, long = Schmidt-Kalman filter}
\DeclareAcronym{LRO}{short = LRO, long = Lunar Reconnaissance Orbiter}
\DeclareAcronym{ETP}{short = ETP, long = Europa Tomography Probe}
\DeclareAcronym{ISL}{short = ISL, long = Inter-Satellite Link}
\makeatletter
\def\ps@pprintTitle{%
 \let\@oddhead\@empty
 \let\@evenhead\@empty
 \def\@oddfoot{}%
 \let\@evenfoot\@oddfoot}
\makeatother
\begin{document}


\begin{frontmatter}

\title{Performance Analysis of Crosslink Radiometric Measurement based Autonomous Orbit Determination for Cislunar Small Satellite Formations}%

\author{Erdem Turan\corref{cor1}}
\cortext[cor1]{Corresponding author}
  \ead{e.turan@tudelft.nl}
\author{Stefano Speretta}
\author{Eberhard Gill}

\address{Delft University of Technology, Faculty of Aerospace Engineering, Kluyverweg 1, 2629 HS, Delft, the Netherlands}

\begin{abstract}
Recent advances in space technology provide an opportunity for small satellites to be launched in cislunar space. However, tracking these small satellites still depends on ground-based operations. Autonomous navigation could be a possible solution considering challenges presented by costly ground operations and limited onboard power available for small satellites. There have been various studies on autonomous navigation methods for cislunar missions. One of them, LiAISON, provides an autonomous orbit determination solution solely using inter-satellite measurements. This study aims at providing a detailed performance analysis of crosslink radiometric measurements based autonomous orbit determination for cislunar small satellite formations considering the effects of measurement type, measurement accuracy, bias, formation geometry and network topology. This study shows that range observations provide better state estimation accuracy than range-rate observations for the autonomous navigation system in cislunar space. Line-of-sight angle measurements derived from radiometric measurements do not improve the overall system performance. In addition, less precise crosslink measurement methods could be an option for formations in the high observable orbital configurations. It was found that measurement biases and measurements with high intervals reduce the overall system performance. In case there are more than two spacecraft in the formation, the navigation system in the mesh topology provided better overall state estimation than centralized topology. 
\end{abstract}

\end{frontmatter}



\section{Introduction}
\label{sec1}

In recent years, there has been a growing attention in small satellite missions to the Moon. A significant interest can be observed in cislunar space due to piggyback launch opportunities and the availability of data relay satellites in lunar orbits. The upcoming launch of the Artemis 1 mission, for example, provides an opportunity for exploring cislunar space with thirteen 6U sized CubeSats \cite{EM12015}. These CubeSats have a variety of unique objectives on the way to the Moon. In addition to the Artemis 1 mission, there are other small satellite missions led by various organizations such as ESA, (\cite{Speretta2018, kruzelecky2018vmmo, Walker2018, tortoradidymos, europaclipper1, burgett2016mini, thelen2017europa, imken2016sylph, etp2}) and various studies proposed in literature (\cite{mercer2018planetary, mercer2019small, halodavid, conigliaro2018design, bentum2018cubesat}). All those small satellite missions consider traditional ground-based navigation techniques, but this approach could be expensive, while the development of these missions is expected at a low cost. In addition, it is difficult to track all these small satellites due to the limited capacity of ground stations. Limitations also come from the satellites, such as onboard power available for communications. Considering all these challenges, an autonomous navigation system for cislunar missions could provide a possible solution.

There have been various studies on autonomous navigation methods for the near-Earth missions (\cite{Rebordao2013, Sheikh2006, Hill2008}). These methods are in general visual-based or based on crosslink radiometric measurements. This study focuses on the investigation of crosslink radiometric navigation method which is a promising method for small satellites due to its simplicity and the use of existing technologies and systems. Cislunar space generally refers to the volume between Earth and the Moon including lunar orbits, orbits around L1 point, near-rectilinear halo orbits, and others. However, L2 Halo orbits are also included into analysis in this study. The crosslink radiometric navigation in cislunar space uses the \ac{LiAISON} method which is an orbit determination method using solely satellite-to-satellite observations, such as range and/or range-rate, to estimate absolute states of the involved spacecrafts when at least one of the satellite orbits has a unique size, shape, and orientation \cite{hillthesis}. The characteristics of the acceleration function determine whether inter-satellite range or range-rate measurements can be used alone to estimate the absolute spacecraft states (position and velocity). In a symmetrical gravity field, there is no unique orbital configurations resulting from the acceleration function, also having a symmetric time derivative, leading to no absolute position determination. However, cislunar missions could benefit from the asymmetric gravity or unique gravity field to build a dedicated positioning system. 

The orbit determination performance of \ac{LiAISON} depends on various factors, such as measurement type, accuracy, bias, frequency, relative geometry between satellites, and others. Regarding the measurement type, range-rate measurements provide better ground-based navigation solutions for deep space or cislunar missions than range measurements. However, for radio frequency based crosslink autonomous navigation applications, a detailed analysis is required on which data types (range, range-rate and \ac{LOS} angles) provide superior navigation solution. On the other hand, if the small satellite uses ranging for its navigation, the ranging signal reduces the power available for telemetry, which reduces the data rate that can be supported. For such cases, various ranging methods have been proposed in literature including time-derived and telemetry-based ranging methods (\cite{andrews2010telemetry, hamkins2015telemetry, openradio2018}). However, these techniques don't provide accurate ranging solutions as using conventional methods and thus this would affect the orbit determination performance. This requires a realistic comparison between data types considering their expected accuracy to choose the best data type for the orbit determination process. In addition to those, other aspects may affect the navigation performances: relative geometry between satellites and number of satellites in the system. These points require a detailed investigation to understand the limits of crosslink radiometric navigation in use of cislunar missions. This study aims at providing a realistic performance analysis. It also presents analytical calculations for the special case of autonomous spacecraft at cislunar orbits in particular by showing the results of observability, covariance, consider-covariance and Monte Carlo analysis.

This paper organized as follows: Section~\ref{sec2} shows the crosslink radiometric navigation method. In Section~\ref{sec3}, corresponding radiometric measurements are presented. Performance analysis tools and results are given in Section~\ref{sec4} and~\ref{sec5}, respectively. Lastly, conclusion is drawn in Section~\ref{sec6}.

\section{Crosslink Radiometric Navigation}
\label{sec2}

In general, small satellites do not have sufficient onboard power for direct-to-Earth communication and data transmission in deep space or in cislunar space. Therefore, mother spacecraft can be involved to support small satellites for such cases. This also allows to use existing systems and technologies for navigation purposes. Basically, relative radiometric measurements, relative range/range-rate/angle, provide a relative navigation solution for the distributed satellite systems. Having an absolute state information for the mothercraft is sufficient to derive the daughtercraft absolute states via the relative navigation solution. However, relative range measurements alone are not sufficient to determine the full states for Earth orbiting satellites due to the rank defect problem \cite{Gao2014}. Even using both range and range-rate measurements does not provide a full state estimation in the two-body problem \cite{meng2010autonomous}.

\begin{figure}[h]
\centering
\caption{CAPSTONE will be the first cislunar CubeSat and will perform spacecraft-to-spacecraft radiometric navigation \cite{capstonenasa}}
\label{capstonefig}
\includegraphics[width=10cm]{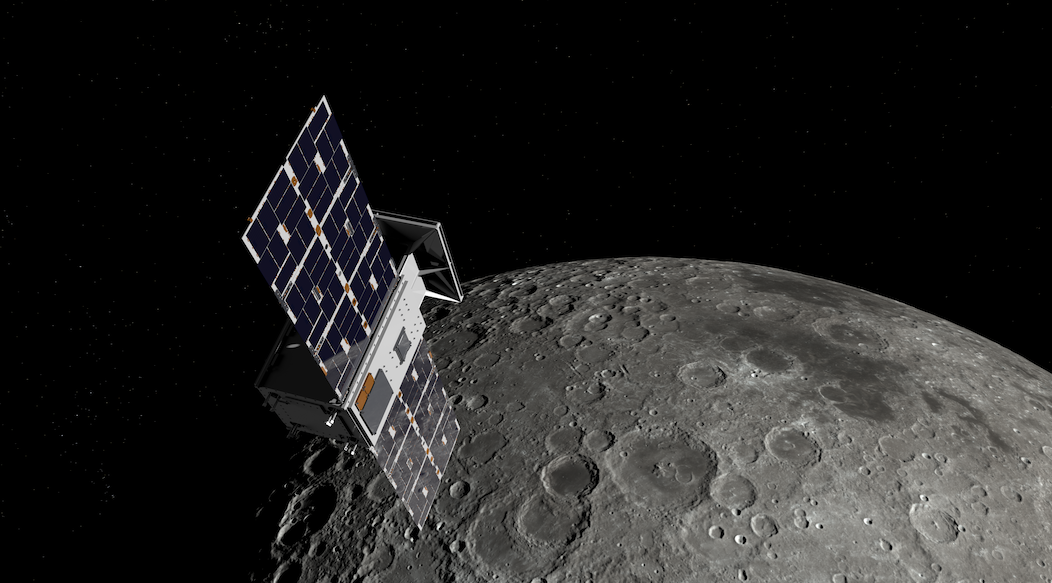}
\end{figure}

Autonomous navigation requires the estimation of absolute position and velocity of a spacecraft without using any ground-based observation. In order to do that, spacecraft states must be observable from the available radiometric measurements between satellites (\cite{fujimoto2012simulating, Hill2007}). In other words, the size, shape, and orientation of the spacecraft orbit must be observable using the available radiometric measurements. Thus, the observability of the system  depends on one of the spacecraft occupying a unique trajectory and this can be used as a reference. In the symmetrical gravity field of the two-body problem such as Earth orbiting satellites, there is no unique orbit due to the acceleration function and its symmetric derivative. Relative measurements, in a two-body problem, do not provide the absolute orientation of the orbital planes but only the relative orientation (\cite{Qin2019, hillthesis}). The \ac{LiAISON} method uses solely inter-satellite measurements to estimate absolute states when at least one of the spacecraft orbit has a unique size, shape, and orientation which can be found in cislunar space e.g. around libration points. Basically, gravitational perturbations of the Moon are sufficient for such unique orbital configurations to exist at the cislunar space and the Earth-Moon Lagrangian points. This method has been applied to mission studies at librations points, asteroids, and cislunar vicinity (\cite{hillthesis, Hill2007, Leonard2012, inproceedingsHesar, Wang2019, Hill2006, Hill2008, leonard2015thesis,  Fujimoto2016}). The method will be applied in the CAPSTONE mission (Figure~\ref{capstonefig}) using a crosslink between \ac{LRO} and the CAPSTONE CubeSat, (\cite{capstonenasa}).

To achieve a better navigation accuracy via the \ac{LiAISON} method, in general, non-coplanar orbital configurations and varying inter-satellite distances are required. The more satellites are added to the constellation and having ground-based observations also increase the overall navigation accuracy. In this paper, these and other misison aspects will be studied with the help of performance analysis tools.

\section{Crosslink Radiometric Data}
\label{sec3}

Inter-satellite radiometric measurements can provide data for autonomous navigation: this section presents three main data types (range, range-rate and \ac{LOS} angles) derived from the radiometric measurements and presents how can they contribute to the autonomous navigation strategy.

\subsection{Range}

The inter-satellite distance can be measured either by the signal travel time from a transmitter (signal source) to a receiver or by the phase shift on a ranging signal at a receiver with respect to a transmitter. This process can either be one-way or two-way. However, one-way tracking requires a very accurate time synchronization between the involved spacecraft. This is very hard to achieve in practice, especially on small satellites with limited capabilities,  so this option is not considered in this study. 

In a conventional two-way ranging using a ground station, an uplink ranging signal is received and re-transmitted by the satellite. The downlink ranging signal is then received at the ground station to allow the computation of the two-way signal traveling time. The ranging signal can either be formed by sequential tones or a pseudonoise code, providing flexibility to the mission designers that can select the best performing solution. This ranging signal retransmission can either be regenerative or transparent, depending on the amount of processing performed on-board the satellite. In regenerative ranging, the spacecraft demodulates and acquires the ranging code by correlation with a local replica from the uplink ranging signal and regenerates the ranging code on the downlink. This allows to reduce the noise influence on the re-transmitted signal, actually lowering the required signal strength at the satellite receiver. In transparent ranging, the spacecraft translates the uplink ranging signal to the downlink without code acquisition \cite{book2014pseudo}. It would also possible to use these techniques for inter-satellite ranging when the full logic could be implemented on a satellite.

Most commonly used radio-navigation transponders generate a downlink carrier that is phase-coherent to the uplink signal to maximize the navigation performances. However, small satellites often lack radio links with coherent operations and thus using non-coherent pseudo-noise ranging which introduces a range bias in the measurements due to a chip rate mismatch between the replica and received codes \cite{book2014pseudo} (besides also decreasing the range-rate estimation accuracy). In case of applying this method to a satellite formation in the cislunar space and considering a non-coherent transponder with a pseudo-noise square-wave shaped ranging signal, a chip tracking loop and an on-board loop bandwidth on the mothercraft wider than daughtercraft's, the following one-way ranging error would be observed~\cite{book2014pseudo}:
\begin{eqnarray}
    \label{rsigma}
    \sigma _{\rho_{PN}} =\frac{c}{8 f_{rc}}\sqrt{\frac{B_{L}}{(P_{RC}/N_{0})}}
\end{eqnarray}

And the range bias due to a chip rate mismatch:
\begin{eqnarray}
    \label{rbias}
    \rho _{\text{bias}}=\frac{c\Delta f_{chip}T}{4f_{chip}}
\end{eqnarray}

with $c$ the speed of light, $f_{rc}$ the frequency of the ranging clock component, $B_{L}$ one-sided loop noise bandwidth, $P_{RC}$ power of the ranging clock component, $T$ integration time, $N_{0}$ one-side noise power spectral density, $\Delta f_{chip}$ the difference in frequency between the received chip rate and the local chip rate. 

However, using conventional ranging methods would require a certain amount of onboard power for the downlink ranging signal. This results in limited power availability for the telemetry signal. This issue could be solved by performing telemetry-based ranging which provides a round-trip light time solution derived from the telemetry stream (\cite{hamkins2015telemetry, andrews2010telemetry}). Telemetry-based ranging does not require any downlink ranging signal but it provides the delay between acquired uplink ranging signal and the start of the next telemetry stream. Basically, all timing data are collected by the signal source to calculate a round-trip light time solution which includes the two-way light time propagation and the re-transmission delay. One of the major advantages of this method is having ranging and telemetry transmission at the same time, removing the need to perform separate tracking sessions or multiplexing tracking and telemetry sessions. Secondly, this method is based on the data-rate and even low data rates would provide a ranging solution as good as conventional tone or pseudo-noise ranging. Considering a direct-to-Earth link, telemetry-based ranging provides better than conventional pseudo-noise ranging measurements at a data rate of about 15~kbps while using a correlator method \cite{andrews2010telemetry}. Considering a square wave uplink range clock and BPSK-modulated data, performance of the telemetry-based ranging can be given as \cite{andrews2010telemetry}:
\begin{eqnarray}
    \label{tmrange}
    \sigma _{\rho_{TM}} = \left ( 1-\frac{2v}{c} \right ) \left (\frac{4 \, c \, T_{sd}^{2}}{\pi \, T_{l} \, E_{S}/N_{0}} + \frac{c}{8 f_{rc}}\sqrt{\frac{B_{L}}{(P_{RC}/N_{0})}} \right )  
\end{eqnarray}

with $T_{sd}$ the channel symbol duration, $T_{l}$ the correlator integration time and $E_{S}/N_{0}$ the code symbol-to-noise ratio. 

In the telemetry-based ranging, the uplink ranging signal is yet required and the traditional ground uplink signals could be used~\cite{hamkins2015telemetry} but any type of uplink signal could be used as well: a more power-efficient solution could be selected, for example, for a small satellite implementing a cross-link. 

Another way to compute the round-trip light time between satellites is based on time transfer, as in the CCSDS Proximity-1 Space Link Protocol \cite{prox1}, where time correction, correlation and distribution are standard services. Users can exchange epochs between satellites and derive the round-trip light time. This process requires obtaining four successive timestamps (time of transmission and reception of both spacecraft) and calculating the round trip light time and offset \cite{Woo2010}.

\begin{figure}[b]
\centering
\includegraphics[width=7cm]{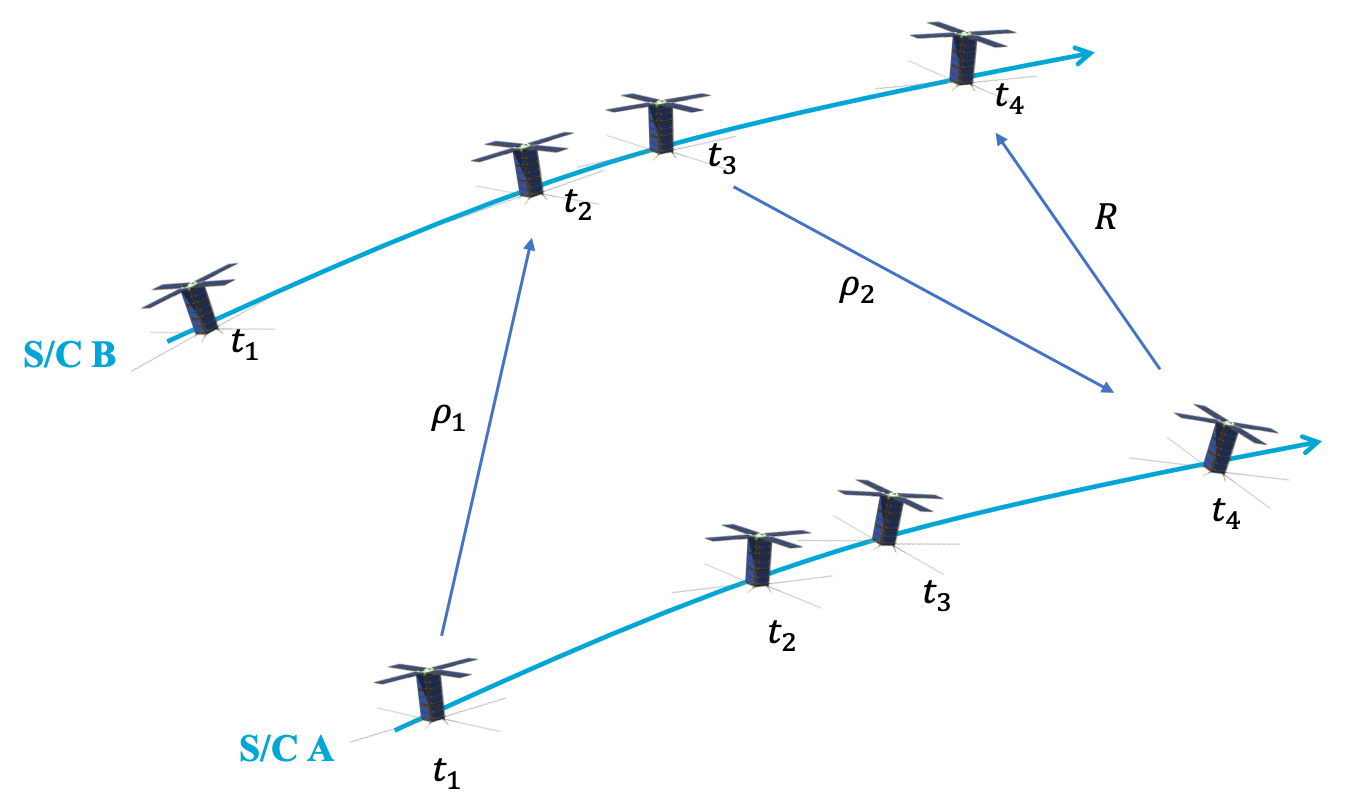}
\caption{Representation of the crosslink range measurement.}
\label{timedis}
\end{figure}

In \cite{openradio2018}, a very similar method was proposed, where the round-trip light time has been measured from ping requests directly using the satellite radio. From the hardware testing, a ranging accuracy has been found as \SI{155}{m} ($1\sigma$) under strong signal conditions and \SI{303}{m} ($1\sigma$) under realistic worst-case conditions for \SI{10}{kbps} data rate. Strong signal conditions refer to a \ac{BER} $10^{-5}$ or lower while worst case conditions refer to a \ac{BER} of $10^{-4}$. Basically, this method does not provide an accurate ranging solution, but this can still be sufficient to meet navigation requirements for certain missions. If timing is measured in units of telemetry/telecommand symbols, instead of directly in seconds, the downlink equation given in Eq.~(\ref{tmrange}) of the telemetry-based ranging could be used for both links. Based on the same assumptions used in Eq.~(\ref{tmrange}), and assuming $T_{l}, \, E_{S}/N_{0}$ are the same on both downlink and uplink sides, the performance of the time-derived ranging is:
\begin{eqnarray}
    \label{sigmaTDrange}
    \sigma _{\rho_{TD}} = \left ( 1-\frac{2v}{c} \right ) \left ( \frac{4 \, c \, \sqrt{T_{sd \, U}^{4}+T_{sd \, D}^{4}}}{\pi \, T_{l} \, E_{S}/N_{0}}\right )
\end{eqnarray}

with $T_{sd \, U}$ and $T_{sd \, D}$ the symbol duration for uplink and downlink respectively. 

\begin{figure}[h]
\centering
\includegraphics[width=8cm]{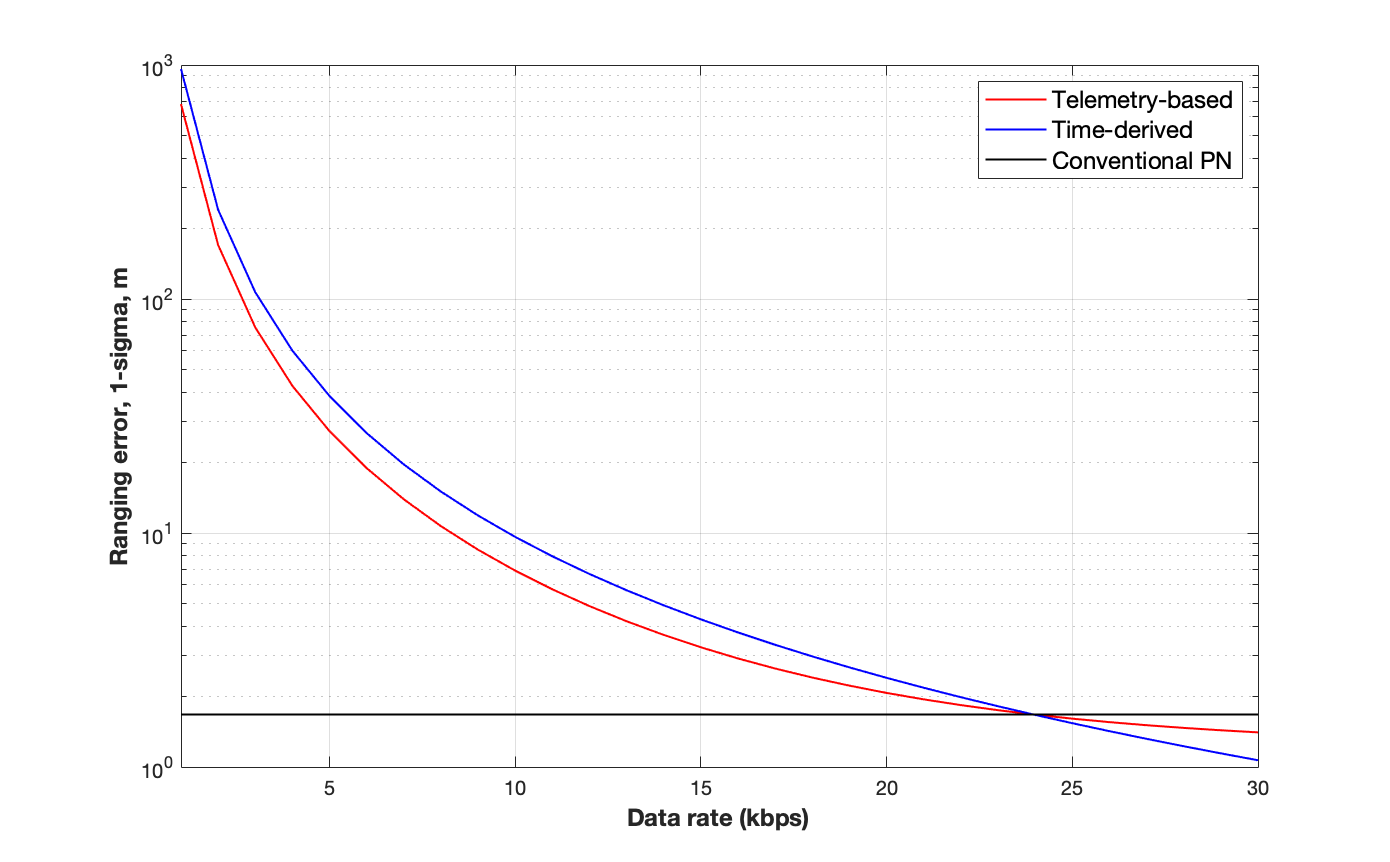}
\caption{Performance comparison of the ranging methods.}
\label{rmethods}
\end{figure}

Comparison of all these three ranging methods is shown in Figure~\ref{rmethods}. As it can be seen, the conventional pseudo-noise ranging is not a function of data-rate (but of the navigation signal chip rate, independent from the data transmission rate). On the other hand, the telemetry-based ranging and the time-derived ranging methods show improved performances with increased data rate.

\subsection{Range-rate}

A basic physical phenomena, the Doppler effect, is used to  measure the relative range rate. Basically, the Doppler shifted transmitted signal frequency arriving at receiver provides an estimation of the relative velocity. It is also possible to derive range-rate from successive range measurements but all these techniques suffer from measurement errors. 

This observation type is also affected by both random (instrumental and propagation noise) and systematic errors \cite{Asmar2005}. However, the most dominant error for Doppler measurement is thermal noise. Measurement error for two-way Doppler due to thermal noise can be approximated by \cite{dsnranging}:
\begin{eqnarray}
    \label{rvsigma}
    \sigma _{V} = \frac{c}{2 \sqrt{2} \pi f_{c} T}\sqrt{\frac{1}{\rho_{L}}+\frac{G^{2} B_{L}}{(P_{C}/N_{0})}}
\end{eqnarray}

where $f_{c}$ the downlink carrier frequency, $P_{C}/N_{0}$ uplink carrier power to noise spectral density ratio, $\rho_{L}$ the downlink carrier loop signal-to-noise ratio, $G$ the turn-around ratio. In one of the Europa Clipper piggyback mission proposals, the \ac{ETP} \cite{etp2} two-way coherent X-band Doppler data from the \ac{ISL} between Europa Clipper and \ac{ETP} will be used for orbit determination (with an expected range rate error of $\SI{0.012}{mm/s}$ with an integration time $\tau$ of $\SI{60}{s}$).

\subsection{Line-of-Sight direction}
\label{LOSsection}

\begin{figure}[h]
\centering
\includegraphics[width=0.33\textheight]{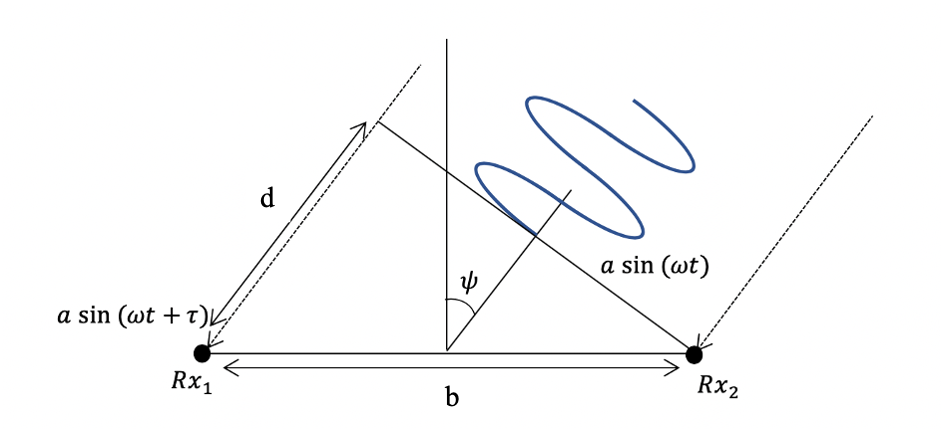}
\caption{\ac{LOS} measurement via two-antennas}
\label{Losfig}
\end{figure}

The full \ac{LOS} direction can be calculated via the time delay or phase shift of the incoming signal with at least three antennas, in order to calculate two angles, mounted on a baseline at a certain distance, $b$, \cite{Fehse2003}. Figure~\ref{Losfig} shows the principle of \ac{LOS} measurement via two antennas. On small satellites, due to size constraints, antenna baseline is often shorter than the wavelength, simplifying the calculations as wavelength ambiguity is eliminated. The following equations can be used to estimate the angle $\psi $ between the line toward the signal source and the line perpendicular to the baseline, $b$, via the time delay $\Delta t$:
\begin{eqnarray}
\label{eqndeltat}
    \Delta t = \frac{b}{c}\,\cos \psi 
\end{eqnarray}

or via the phase-shift,
\begin{eqnarray}
\label{eqntau}
    \tau = \frac{2\pi b}{\lambda}\,\sin \psi 
\end{eqnarray}

Basically, the \ac{LOS} measurement accuracy depends on length of the baseline and time-delay/phase-shift measurement accuracy (ranging accuracy). Taking the variance of both hand sides of the above equations would give the relation between range measurement accuracy $\sigma_\rho$ and the \ac{LOS} measurement accuracy $\sigma_{\psi}$, via the time delay (See \ref{appendixa} for the full derivation):
\begin{eqnarray}
\label{eqntimedelay}
    \sigma_{\rho}= b \sqrt{\frac{1-e^{-2\sigma_{\psi}^{2}}}{2}}
\end{eqnarray}

or via the phase-shift:
\begin{eqnarray}
    \sigma_\tau = \frac{2\pi b}{\lambda} \sqrt{\frac{1-e^{-2\sigma_{\psi}^{2}}}{2}}
\end{eqnarray}

\begin{figure}[h]
\centering
\includegraphics[width=0.33\textheight]{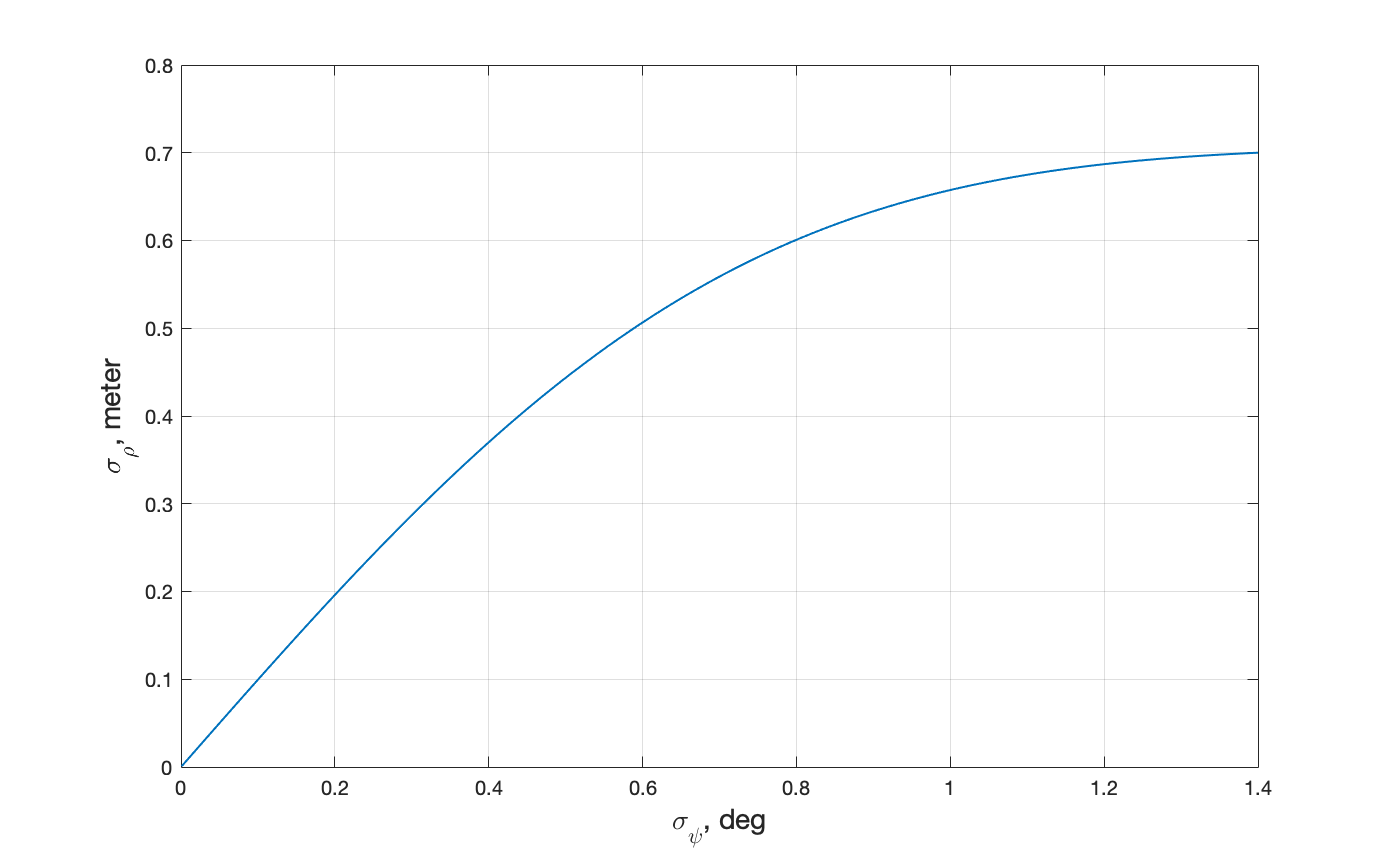}
\caption{Relation between \ac{LOS} angle and range measurement errors}
\label{Loserrorfig}
\end{figure}

The relation between \ac{LOS} and the range measurement error can be seen in Figure~\ref{Loserrorfig} for a baseline of \SI{1}{m}. 

\section{Orbit Determination Models}
\label{sec4}

In this section, orbit determination models used in this study are presented. Dynamical model, observation models and estimation models are shown in the following subsections.

\subsection{Orbit Dynamics Model}

\begin{figure}[h]
\centering
\includegraphics[width=0.33\textheight]{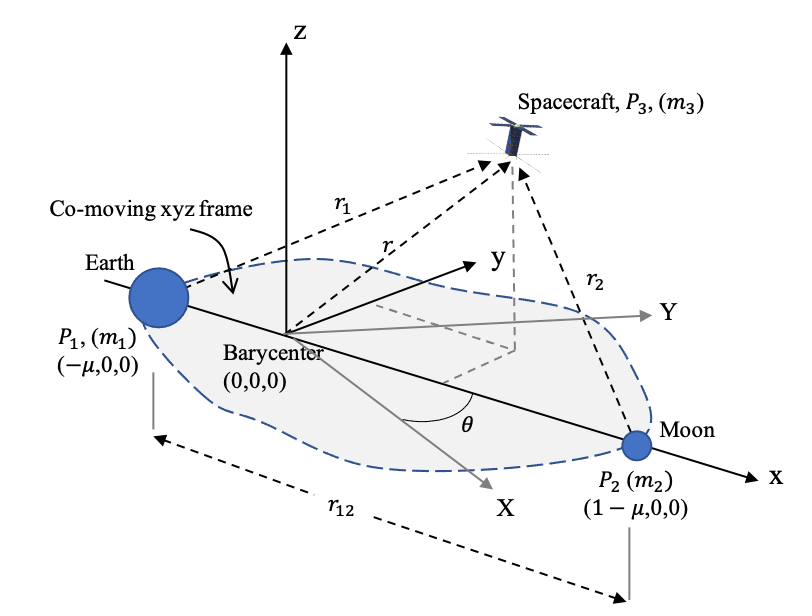}
\caption{Formulation of the Circular Restricted Three-Body Problem: a rotating, non-dimensional coordinate frame.}
\label{fig:CRTBPf}
\end{figure}

This section presents the orbital dynamics model used in this paper. Dynamic models are formulated based on the \ac{CRTBP}. This model is simple but accurate enough for many applications: in \cite{hillthesis}, the autonomous orbit determination accuracy remained at the same order of magnitude for various force models (\ac{CRTBP} and JPL DE405 Ephemeris). Therefore, the \ac{CRTBP} is selected for analysis in this study. The \ac{CRTBP} assumes that there are two massive bodies, Earth ($P_1$) and Moon ($P_2$) in this case with masses $m_1$ and $m_2$ respectively, moving under their mutual gravitation in a circular orbit around each other with a radius $r_{12}$, \cite{Curtis2020}. Considering a non-inertial, co-moving reference frame (Figure~\ref{fig:CRTBPf}) whit its origin at the barycenter of the two bodies, the $x-$direction pointing from barycenter to $P_2$. The positive $y-$axis is parallel to the velocity vector of $P_2$ and $z-$axis is perpendicular to the orbital plane. 

Considering a third body of mass $m_3$ with $m_3 \ll m_1$ and $m_3 \ll m_2$, it cannot impact the motion of primary bodies, $P_1$ and $P_2$. Using Newton's second law, the equations describing the motion of the third body $P_3$
\begin{eqnarray}
\label{eqnnewtonsec}
    m_3\ddot{\mathbf{r}}=-\frac{Gm_1 m_3}{r_1^{ 3}}\mathbf{r}_1-\frac{Gm_2 m_3}{r_2^{3}}\mathbf{r}_2
\end{eqnarray}

where $G$ is the universal gravitational constant. Note that the absolute acceleration of $m_3$ requires differentiation in the inertial frame \cite{daniel2006generating}. By using the characteristic quantities
\begin{eqnarray}
    l^*=r_{12}, \hspace{5mm} t^*=\sqrt{l^{*^{3}}/G(m_1 + m_2)}
\end{eqnarray}

Eq.~(\ref{eqnnewtonsec}) can be non-dimensionalize as
\begin{eqnarray}
\label{eqnnewtonsecnondim}
    \frac{\mathrm{d}^2 \mathbf{r}}{\mathrm{d} \tau^2}=-\frac{\mu}{r_1^{3}}\mathbf{r}_1-\frac{1-\mu}{r_2^{3}}\mathbf{r}_2
\end{eqnarray}

where $\mu=m_2 / (m_1 + m_2)$ and the non-dimensional time $\tau=t/t^*$. Note that the position vectors, $\mathbf{r},\mathbf{r}_1$ and $\mathbf{r}_2$, are now non-dimensional (e.g., $\mathbf{r}_{\text{non-dim}}=\mathbf{r}_{\text{dim}}/l^*$). The angular velocity of the rotating frame with respect to the inertial frame is written $\mathbf{\Omega} = n \hat{z}$ where the non-dimensional mean motion is $n=1$, \cite{daniel2006generating}. After the differentiation of $\mathbf{r}$ twice with respect to an inertial observer, the equations of motion for the \ac{CRTBP} \cite{daniel2006generating} are
\begin{eqnarray}
\ddot{x}-2\dot{y}=x-(1-\mu)\frac{x+\mu}{r_1^3}-\mu\frac{x+\mu-1}{r_2^3}
\end{eqnarray}
\begin{eqnarray}
    \ddot{y}+2\dot{x}=(1-\frac{1-\mu}{r_1^3}-\frac{\mu}{r_2^3})y 
\end{eqnarray}
\begin{eqnarray}
\ddot{z}=(\frac{\mu-1}{r_1^3}-\frac{\mu}{r_2^3})z
\end{eqnarray}

where $r_1=\sqrt{(x+\mu)^2 + y^2 + z^2}$ and $r_2=\sqrt{(x+\mu-1)^2 + y^2 + z^2}$. For the Earth-Moon system, $\mu = 0.01215$, $t^* = 4.343$, and $l^* = 384747.96$, respectively.

\subsection{Observation Models}
Considering a formation formed by two spacecraft, the estimated state vector with position and velocity components of both spacecraft is
\begin{eqnarray}
\mathbf{X}=\begin{bmatrix}
x_1 & y_1 & z_1 & {\dot{x}}_1 & {\dot{y}}_1 & {\dot{z}}_1 & x_2 & y_2 & z_2 & {\dot{x}}_2 & {\dot{y}}_2 & {\dot{z}}_2
\end{bmatrix}^T
\end{eqnarray}

Measuring the round-trip light time in general is based on phase measurement of a ranging signal and, in this way, the on-board clock will be used as time reference, potentially causing a measurement bias. This bias either measured along with the navigation filter or calibrated. The measurement model in this paper, referred as the pseudorange, involves the geometric range, the overall clock bias, and other error sources. In the Figure~\ref{timedis} and \ref{2wrtlt}, the concept of the two-way ranging measurement can be seen. Basically, spacecraft A transmits a ranging signal at time $t_1$ to spacecraft B (who receives it at time $t_2$ and re-transmits it at time $t_3$)  and receives it back at time $t_4$. During this measurement interval, both spacecraft are moving in their orbits, so there are also changes in line-of-sight direction which can be modeled as $\Delta \rho$. In the end we can model the geometric range as: 

\begin{figure}[h]
\centering
\includegraphics[width=0.33\textheight]{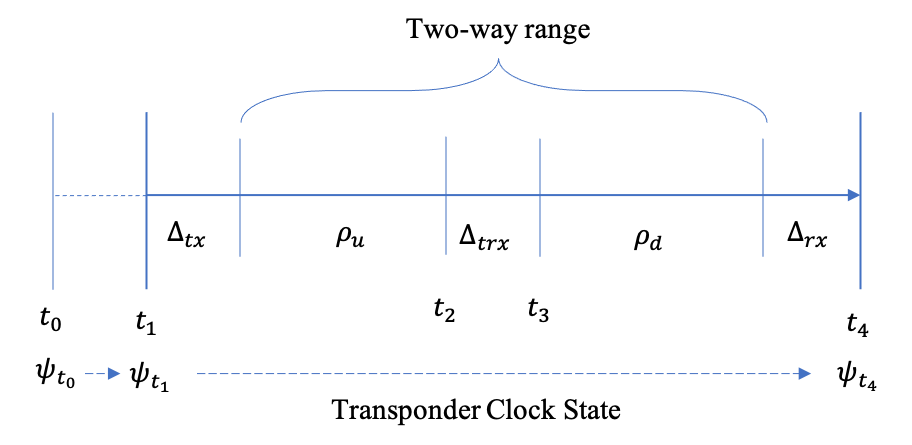}
\caption{Two-way round-trip light time measurement.}
\label{2wrtlt}
\end{figure}
\begin{eqnarray}
    R=\frac{1}{2}\,c\,(t_4 - t_1 ) + \Delta \rho
\end{eqnarray}

By assuming the speed of light is greater than the spacecraft relative velocity, $c\gg v$, and ignoring the light-time correction, the geometric range can be modeled as:
\begin{eqnarray}
    R= \sqrt{(x_1 -x_2)^{2}+(y_1 -y_2)^{2}+(z_1 -z_2)^{2}}
\end{eqnarray}

Then the pseudorange observations can be modeled as:
\begin{eqnarray*}
    \rho = R + c(\psi _{t_3}-\psi _{t_1})+c\,(\Delta_{tx}+\Delta_{rx})+c\,\Delta_{trx}+\rho_{\text{noise}}
\end{eqnarray*}
\begin{eqnarray}
    \rho = \sqrt{(\mathbf{r}_1-\mathbf{r}_2)\cdot (\mathbf{r}_1-\mathbf{r}_2)}+\rho_{\text{bias}} + \rho_{\text{noise}}
\end{eqnarray}

where $\psi _{t_4}$ and $\psi _{t_1}$ are the clock states at $t_4$ and $t_1$ respectively. The transponder transmit and receive line delays are $\Delta_{tx}$ and $\Delta_{rx}$, respectively and $\Delta_{trx}$ is the line delay on the spacecraft transponding the ranging signal. All these terms are combined as $\rho_{\text{bias}}$ and $\rho_{\text{noise}}$ representing the unmodelled statistical error sources.

The range rate measurements, $\dot{\rho}$, can be modelled as:
\begin{eqnarray}
    \dot{\rho}=\frac{\boldsymbol{\rho}\cdot\dot{\boldsymbol{\rho}}}{\rho}
\end{eqnarray}
\begin{eqnarray}
    \dot{\rho}=\frac{\left(x_1-x_2\right)\left({\dot{x}}_1-{\dot{x}}_2\right)+\ \left(y_1-y_2\right)\left({\dot{y}}_1-{\dot{y}}_2\right)+\left(z_1-z_2\right)\left({\dot{z}}_1-{\dot{z}}_2\right)}{\sqrt{\left(x_1-x_2\right)^2+\left(y_1-y_2\right)^2+\left(z_1-z_2\right)^2}} + \dot{\rho}_{\text{bias}} + \dot{\rho}_{\text{noise}}
\end{eqnarray}

Finally, the \ac{LOS} angle measurements can be modeled as:
\begin{eqnarray}
    \phi=\arctan \left ( \frac{y_2 -y_1}{x_2 -x_1} \right ) + \phi_{\text{bias}} + \phi_{\text{noise}}
\end{eqnarray}
\begin{eqnarray}
    \varphi=\arcsin \left ( \frac{z_2 -z_1}{\sqrt{(x_1 -x_2)^{2}+(y_1 -y_2)^{2}+(z_1 -z_2)^{2}}} \right ) + \varphi_{\text{bias}} + \varphi_{\text{noise}}
\end{eqnarray}

where $\phi$ is a relative azimuth angle, $\varphi$ is a relative elevation angle between the two spacecraft. 

\subsection{Estimation Model}

In this study, an \ac{EKF} is adopted as a common method used in real-time navigation. The \ac{EKF} consists of a prediction step and a correction step. In the prediction step, predicted state and error covariance prediction $\bar{P}$ are \cite{Tapley2004}
\begin{eqnarray}
    \dot{\mathbf{X}} = \boldsymbol{F}({\mathbf{X}},t), \hspace{5mm} {\mathbf{X}}(t_{k-1})=\hat{\mathbf{X}}_{k-1}
\end{eqnarray}
\begin{eqnarray}
\label{eqnPmeas}
\boldsymbol{\bar{P}}_k = \boldsymbol{\Phi}(t_k,t_{k-1})\boldsymbol{P}_{k-1}\boldsymbol{\Phi}^T(t_k,t_{k-1}) + \boldsymbol{Q}
\end{eqnarray}

where $\boldsymbol{\Phi}(t_k,t_{k-1})$ is state transition matrix from $t_{k-1}$ to $t_{k}$ and $\boldsymbol{Q}$ is the process noise matrix. The correction step:
\begin{eqnarray}
    \boldsymbol{K}_k=\boldsymbol{\bar{P}}_k \boldsymbol{\tilde H}_k^T [\boldsymbol{\tilde H}_k \boldsymbol{\bar{P}}_k \boldsymbol{\tilde H}_k^T +\boldsymbol{W}_k]^{-1}
\end{eqnarray}
\begin{eqnarray}
    \hat{\mathbf{X}}_k = {\mathbf{X}}_k + \boldsymbol{K}_k [\mathbf{y}_k - \boldsymbol{\tilde H}_k {\mathbf{X}}_k ]
\end{eqnarray}
\begin{eqnarray}
\label{eqnPtime}
    \boldsymbol{P}_k =[\boldsymbol{I}-\boldsymbol{K}_k \boldsymbol{\tilde H}_k] \boldsymbol{\bar{P}}_k
\end{eqnarray}

where $\bold{\hat{X}}$ is state estimate, $\boldsymbol{K}$ is Kalman gain, $\boldsymbol{\tilde H}$ is the measurement sensitivity, $\boldsymbol{P}$ is error covariance estimate, and $\boldsymbol{W}$ is the state noise compensation matrix. 

\section{Performance Analysis}
\label{sec5}

Throughout this study, several performance analysis methods were used, including the observability analysis, the covariance analysis, and the Monte Carlo analysis. This section begins with the basic concept of observability to present its definition and the observability criteria for crosslink radiometric measurement based autonomous navigation, mainly introducing the observability analysis methods. As part of the observability analysis, \acf{SVD}, observability Gramian, and estimation error covariance were considered. Next,  the Monte Carlo analysis is discussed. Lastly, an observation type comparison is provided for the autonomous navigation system, showing which observable (range, range-rate and \ac{LOS}) would provide better results.

\subsection{Observability Analysis}
\label{subsecobservability}

The observability analysis of an autonomous navigation system allows to investigate the relation between navigation accuracy and measurement type, frequency, and accuracy. This analysis also allows to investigate the relation between the estimation accuracy and other aspects such as inter-satellite distance, orbital periods, and number of spacecraft. Basically, the observability analysis may help testing whether the navigation parameters could be determined with observation data, but it cannot directly evaluate the estimation accuracy. Actually, the degree of observability can be used alone to evaluate the estimation accuracy. If the degree of observability or in other words state observability is increased, this means an increased reduction of estimated state uncertainty \cite{Dianetti2017}. The standard approach to measure observability is checking the observability rank condition which, unfortunately, only provides information whether the system is observable or not. The degree of observability can be either checked via the observability Gramian or the information matrix. Both methods are related to each other via the measurement covariance matrix which provides weighted information about its observability. One of the possible observability assessment would be the use of \ac{SVD} of either the observability Gramian or the information matrix. This provides the observability of the modes of the system allowing to find higher and lower observable states. It is possible to evaluate the degree of observability by checking the condition number, which is the ratio of the largest singular value to the smallest one. Another possible measure is the local unobservability index which is the reciprocal of the smallest local singular value. If it is large, then observation noise would have a large impact on the estimation error \cite{Krener2009}. The local unobservability index will not increase with additional observations but the condition number might increase. In brief, a higher degree of observability indicates, in general, a more accurate estimation.

Under a discrete time assumption, the time varying observability Gramian can be written as:

\begin{eqnarray}
\label{Neqn}
    \boldsymbol{N}=\sum_{k=1}^{l}\boldsymbol{\Phi}(t_k,t_0)^T \boldsymbol{\tilde{H}}^T_k\boldsymbol{\tilde{H}}_k \boldsymbol{\Phi}(t_k,t_0) 
\end{eqnarray}

A theoretical analysis can be conducted based on the observability Gramian to show the observable states or states combinations. If the Gramian has a full rank, which means all columns are linearly independent from each other, the whole system is observable and all states can be estimated by a navigation filter. If there are linearly dependent columns, the corresponding states are unobservable. If there exists a linear combination of the dependent columns, which is independent from other columns, the corresponding linear state combination is observable, and the total amount of observable states and state combinations is equal to the rank of the observability matrix.

Another observability approach would be based on the information matrix given as:
\begin{eqnarray}
\label{infomatrix}
    \boldsymbol{\Lambda} = \boldsymbol{{H}}^{T} \boldsymbol{W} \boldsymbol{{H}} = \sum_{k=1}^{l}\boldsymbol{{H}}_k^{T} \boldsymbol{W} \boldsymbol{{H}}_k 
\end{eqnarray}

The information matrix is similar to the observability Gramian with the addition of a measurement covariance matrix, $\boldsymbol{W}$. The condition number $cond(\boldsymbol{\Lambda})$ or $cond(\boldsymbol{N})$, which is the ratio of the largest singular value to the smallest one, provides an indication of the observability. Using the information matrix in Equation~\ref{infomatrix}, the \ac{SVD} can be performed as:
\begin{eqnarray}
\label{infom}
    \boldsymbol{\Lambda} = \textbf{U} \boldsymbol{\Sigma} \textbf{V}^{T}
\end{eqnarray}

where \textbf{U} and \textbf{V} are unitary matrices and $\boldsymbol{\Sigma}$ is a matrix of singular values. It is also possible to check the degree of observability by looking at the singular values in a matrix \cite{Yim2000}. The state with the largest singular value is the most observable one and it gives the most observable information. Similarly, the state with the smallest singular value is the least observable in the system. By using $\textbf{V}^{T}$, states corresponding to each singular value can be found: the largest number in the columns of $\textbf{V}^{T}$ corresponding to any singular value shows the related states. By looking at all the columns of $\textbf{V}^{T}$, we can sort the states from the most observable one to the least observable one. In addition to the previous points, it is also possible to check the effectiveness of the observation at $t_i$ alone on position/velocity of a specific spacecraft. This can be done by taking the square root of the maximum eigenvalue of each $\boldsymbol{\Lambda}_k$ or $3\times 3$ position/velocity component of $\boldsymbol{\Lambda}_i$, $(\sqrt{\text{max} \, \text{eig}(\boldsymbol{\Lambda}_{i (3 \times 3)})})$ \cite{hillthesis}. Both observability analysis methods require $\boldsymbol{H}$ and thus $\boldsymbol{\tilde H}$ and $\boldsymbol{\Phi}(t_k,t_0)$ to be calculated. 

The observations can be related to the states with a partial differential matrix at time $t_k$. In case of crosslink range measurements, $\rho$, is the observation type then $\boldsymbol{\tilde H}$ is of the form:
\begin{eqnarray}
    \boldsymbol{\tilde H}_{k}^{\rho}=\begin{bmatrix}
\frac{\partial\rho}{\partial x_1} & \frac{\partial\rho}{\partial y_1} & \frac{\partial\rho}{\partial z_1} & \frac{\partial\rho}{\partial{\dot{x}}_1} & \frac{\partial\rho}{\partial{\dot{y}}_1} & \frac{\partial\rho}{\partial{\dot{z}}_1} & \frac{\partial\rho}{\partial x_2} & \frac{\partial\rho}{\partial y_2} & \frac{\partial\rho}{\partial z_2} & \frac{\partial\rho}{\partial{\dot{x}}_2} & \frac{\partial\rho}{\partial y_2} & \frac{\partial\rho}{\partial{\dot{z}}_2}
\end{bmatrix}
\end{eqnarray}
\begin{eqnarray}
    \boldsymbol{\tilde H}_{k}^{\rho}=\begin{bmatrix} \frac{\left(x_1-x_2\right)}{\rho}&\frac{\left(y_1-y_2\right)}{\rho}&\frac{\left(z_1-z_2\right)}{\rho}&0& 0&0& \frac{-\left(x_1-x_2\right)}{\rho}&\frac{-\left(y_1-y_2\right)}{\rho}& \frac{-\left(z_1-z_2\right)}{\rho}&0& 0&0\end{bmatrix}
\end{eqnarray}

In case of range-rate measurements, $\dot{\rho}$, is the observation type then $\boldsymbol{\tilde H}$ is of the form:
\begin{eqnarray}
    \boldsymbol{\tilde H}_{k}^{\dot{\rho}}=\begin{bmatrix}
\frac{\partial \dot{\rho}}{\partial x_1} & \frac{\partial \dot{\rho}}{\partial y_1} & \frac{\partial \dot{\rho}}{\partial z_1} & \frac{\partial \dot{\rho}}{\partial{\dot{x}}_1} & \frac{\partial \dot{\rho}}{\partial{\dot{y}}_1} & \frac{\partial \dot{\rho}}{\partial{\dot{z}}_1} & \frac{\partial \dot{\rho}}{\partial x_2} & \frac{\partial \dot{\rho}}{\partial y_2} & \frac{\partial \dot{\rho}}{\partial z_2} & \frac{\partial \dot{\rho}}{\partial{\dot{x}}_2} & \frac{\partial \dot{\rho}}{\partial y_2} & \frac{\partial \dot{\rho}}{\partial{\dot{z}}_2}
\end{bmatrix}
\end{eqnarray}
\begin{eqnarray}
    \boldsymbol{\tilde H}_{k}^{\dot{\rho}}=\left[\begin{matrix}\frac{\left({\dot{x}}_1-{\dot{x}}_2\right)}{\rho}-\frac{\left(x_1-x_2\right)\left(\left(x_1-x_2\right)\left({\dot{x}}_1-{\dot{x}}_2\right)+\ \left(y_1-y_2\right)\left({\dot{y}}_1-{\dot{y}}_2\right)+\left(z_1-z_2\right)\left({\dot{z}}_1-{\dot{z}}_2\right)\right)}{\rho^3}\\\frac{\left({\dot{y}}_1-{\dot{y}}_2\right)}{\rho}-\frac{\left(y_1-y_2\right)\left(\left(x_1-x_2\right)\left({\dot{x}}_1-{\dot{x}}_2\right)+\ \left(y_1-y_2\right)\left({\dot{y}}_1-{\dot{y}}_2\right)+\left(z_1-z_2\right)\left({\dot{z}}_1-{\dot{z}}_2\right)\right)}{\rho^3}\\\begin{matrix}\frac{\left({\dot{z}}_1-{\dot{z}}_2\right)}{\rho}-\frac{\left(z_1-z_2\right)\left(\left(x_1-x_2\right)\left({\dot{x}}_1-{\dot{x}}_2\right)+\ \left(y_1-y_2\right)\left({\dot{y}}_1-{\dot{y}}_2\right)+\left(z_1-z_2\right)\left({\dot{z}}_1-{\dot{z}}_2\right)\right)}{\rho^3}\\\frac{\left(x_1-x_2\right)}{\rho}\\\begin{matrix}\frac{\left(y_1-y_2\right)}{\rho}\\\frac{\left(z_1-z_2\right)}{\rho}\\\begin{matrix}-\frac{\left({\dot{x}}_1-{\dot{x}}_2\right)}{\rho}+\frac{\left(x_1-x_2\right)\left(\left(x_1-x_2\right)\left({\dot{x}}_1-{\dot{x}}_2\right)+\ \left(y_1-y_2\right)\left({\dot{y}}_1-{\dot{y}}_2\right)+\left(z_1-z_2\right)\left({\dot{z}}_1-{\dot{z}}_2\right)\right)}{\rho^3}\\-\frac{\left({\dot{y}}_1-{\dot{y}}_2\right)}{\rho}+\frac{\left(y_1-y_2\right)\left(\left(x_1-x_2\right)\left({\dot{x}}_1-{\dot{x}}_2\right)+\ \left(y_1-y_2\right)\left({\dot{y}}_1-{\dot{y}}_2\right)+\left(z_1-z_2\right)\left({\dot{z}}_1-{\dot{z}}_2\right)\right)}{\rho^3}\\\begin{matrix}-\frac{\left({\dot{z}}_1-{\dot{z}}_2\right)}{\rho}+\frac{\left(z_1-z_2\right)\left(\left(x_1-x_2\right)\left({\dot{x}}_1-{\dot{x}}_2\right)+\ \left(y_1-y_2\right)\left({\dot{y}}_1-{\dot{y}}_2\right)+\left(z_1-z_2\right)\left({\dot{z}}_1-{\dot{z}}_2\right)\right)}{\rho^3}\\-\frac{\left(x_1-x_2\right)}{\rho}\\\begin{matrix}-\frac{\left(y_1-y_2\right)}{\rho}\\-\frac{\left(z_1-z_2\right)}{\rho}\\\end{matrix}\\\end{matrix}\\\end{matrix}\\\end{matrix}\\\end{matrix}\\\end{matrix}\right]^T
\end{eqnarray}

If \ac{LOS} angles, $\phi$ and $\varphi$ , are selected observation types, then $\boldsymbol{\tilde H}$ is of the form:
\begin{eqnarray}
    \boldsymbol{\tilde H}_{k}^{\phi}=\begin{bmatrix}
\frac{\partial\phi}{\partial x_1} & \frac{\partial\phi}{\partial y_1} & \frac{\partial\phi}{\partial z_1} & \frac{\partial\phi}{\partial{\dot{x}}_1} & \frac{\partial\phi}{\partial{\dot{y}}_1} & \frac{\partial\phi}{\partial{\dot{z}}_1} & \frac{\partial\phi}{\partial x_2} & \frac{\partial\phi}{\partial y_2} & \frac{\partial\phi}{\partial z_2} & \frac{\partial\phi}{\partial{\dot{x}}_2} & \frac{\partial\phi}{\partial y_2} & \frac{\partial\phi}{\partial{\dot{z}}_2}
\end{bmatrix}
\end{eqnarray}
\begin{eqnarray}
\boldsymbol{\tilde H}_{k}^{\phi}=\begin{bmatrix}
\frac{-(y_1 -y_2)}{(x_1 - x_2)^2((\frac{(y_1 -y_2)^2}{(x_1 -x_2)^2})+1)}\hspace{1mm} \frac{1}{(x_1 -x_2)(\frac{(y_1 - y_2)^2)}{(x_1 - x_2)^2}+1)}\hspace{1mm}0\hspace{1mm}0\hspace{1mm}0\hspace{1mm}0\hspace{1mm}\frac{(y_1 -y_2)}{(x_1 - x_2)^2((\frac{(y_1 -y_2)^2}{(x_1 -x_2)^2})+1)}\hspace{1mm} \frac{-1}{(x_1 -x_2)(\frac{(y_1 - y_2)^2}{(x_1 - x_2)^2}+1)}\hspace{1mm}0\hspace{1mm}0\hspace{1mm}0\hspace{1mm}0
\end{bmatrix}
\end{eqnarray}
\begin{eqnarray}
    \boldsymbol{\tilde H}_{k}^{\varphi}=\begin{bmatrix}
\frac{\partial\varphi}{\partial x_1} & \frac{\partial\varphi}{\partial y_1} & \frac{\partial\varphi}{\partial z_1} & \frac{\partial\varphi}{\partial{\dot{x}}_1} & \frac{\partial\varphi}{\partial{\dot{y}}_1} & \frac{\partial\varphi}{\partial{\dot{z}}_1} & \frac{\partial\varphi}{\partial x_2} & \frac{\partial\varphi}{\partial y_2} & \frac{\partial\varphi}{\partial z_2} & \frac{\partial\varphi}{\partial{\dot{x}}_2} & \frac{\partial\varphi}{\partial y_2} & \frac{\partial\varphi}{\partial{\dot{z}}_2}
\end{bmatrix}
\end{eqnarray}
\begin{eqnarray}
    \boldsymbol{\tilde H}_{k}^{\varphi}=\begin{bmatrix}
\frac{(2x_1 -2x_2)(z_1 -z_2)}{(2(1 - (z_1 - z_2)^2 / ((x_1 - x_2)^2 + (y_1 - y_2)^2 + (z_1 - z_2)^2))^{1/2} ((x_1 - x_2)^2 + (y_1 - y_2)^2 + (z_1 - z_2)^2)^{3/2})}
\\
\frac{(2y_1 -2y_2)(z_1 -z_2)}{(2(1 - (z_1 - z_2)^2 / ((x_1 - x_2)^2 + (y_1 - y_2)^2 + (z_1 - z_2)^2))^{1/2} ((x_1 - x_2)^2 + (y_1 - y_2)^2 + (z_1 - z_2)^2)^{3/2})}
\\ 
\frac{-(x_1^2 - 2x_1x_2 + x_2^2 + y_1^2 - 2y_1y_2 + y_2^2)}{(1 - (z_1 - z_2)^2 / ((x_1 - x_2)^2 + (y_1 - y_2)^2 + (z_1 - z_2)^2))^{1/2} ((x_1 - x_2)^2 + (y_1 - y_2)^2 + (z_1 - z_2)^2)^{3/2}}
\\ 
0
\\
0 
\\
0 
\\ 
\frac{-(2x_1 -2x_2)(z_1 -z_2)}{(2(1 - (z_1 - z_2)^2 / ((x_1 - x_2)^2 + (y_1 - y_2)^2 + (z_1 - z_2)^2))^{1/2} ((x_1 - x_2)^2 + (y_1 - y_2)^2 + (z_1 - z_2)^2)^{3/2})}
\\ 
\frac{-(2y_1 -2y_2)(z_1 -z_2)}{(2(1 - (z_1 - z_2)^2 / ((x_1 - x_2)^2 + (y_1 - y_2)^2 + (z_1 - z_2)^2))^{1/2} ((x_1 - x_2)^2 + (y_1 - y_2)^2 + (z_1 - z_2)^2)^{3/2})}
\\ 
\frac{(x_1^2 - 2x_1x_2 + x_2^2 + y_1^2 - 2y_1y_2 + y_2^2)}{(1 - (z_1 - z_2)^2 / ((x_1 - x_2)^2 + (y_1 - y_2)^2 + (z_1 - z_2)^2))^{1/2} ((x_1 - x_2)^2 + (y_1 - y_2)^2 + (z_1 - z_2)^2)^{3/2}}
\\ 
0
\\ 
0
\\ 
0
\end{bmatrix}^T
\end{eqnarray}

The $\boldsymbol{\tilde H}$ matrix is mapped to the initial epoch $t_0$ through the \ac{STM} as:
\begin{eqnarray}
\label{Hmtx}
    \boldsymbol{{H}}_k = \boldsymbol{\tilde{H}}_k \boldsymbol{\Phi}(t_k,t_0)
\end{eqnarray}

Where $\boldsymbol{\Phi}(t_k,t_0)$ is the \ac{STM} from $t_0$ to $t_k$. The differential equation of the \ac{STM} is given by:
\begin{eqnarray}
\label{markleyeqn}
        \boldsymbol{\dot{\Phi}}(t_k,t_0)= \boldsymbol{A}(\text{t})\boldsymbol{\Phi}(\textit{t}_k,\textit{t}_\textit{0}) = \frac{\partial \boldsymbol{F}( \boldsymbol{X}^*,t)}{\partial \boldsymbol{X}} \boldsymbol{\Phi}(\textit{t}_k,\textit{t}_\textit{0})=\begin{bmatrix}
0_{3 \times 3} & \bold{I}_{3 \times 3} \\ 
\mathbf{G}(t) & 2\mathbf{\Omega}
\end{bmatrix}\boldsymbol{\Phi}(t_k,t_0)
\end{eqnarray}

where 
\begin{eqnarray}
  \mathbf{G}(t)=\begin{bmatrix}
\frac{\partial{\ddot{x}}}{\partial x} &\frac{\partial{\ddot{x}}}{\partial y}  &\frac{\partial{\ddot{x}}}{\partial z} \\ 
\frac{\partial{\ddot{y}}}{\partial x} &\frac{\partial{\ddot{y}}}{\partial y}  &\frac{\partial{\ddot{y}}}{\partial z}\\ 
\frac{\partial{\ddot{z}}}{\partial x} &\frac{\partial{\ddot{z}}}{\partial y}  &\frac{\partial{\ddot{z}}}{\partial z}\\ 
\end{bmatrix}
\end{eqnarray}
\begin{eqnarray}
  \mathbf{\Omega}=\begin{bmatrix}
0 & 1 & 0 \\ 
-1 & 0 & 0 \\ 
0 & 0 & 0
\end{bmatrix}
\end{eqnarray}
\begin{eqnarray}
  \boldsymbol{\Phi}(t_k,t_0) =  \begin{bmatrix}
\frac{\partial \mathbf{r}}{\partial \mathbf{r}_0}  & \frac{\partial \mathbf{r}}{\partial \mathbf{v}_0}\\ 
\frac{\partial \mathbf{v}}{\partial \mathbf{r}_0} & \frac{\partial \mathbf{v}}{\partial \mathbf{v}_0}
\end{bmatrix}
\end{eqnarray}
\begin{eqnarray}
    \boldsymbol{\Phi}(t_0,t_0)=\boldsymbol{I}_{\text{12x12}}
\end{eqnarray}

In general, Eq.\ref{markleyeqn} is computed numerically. On the other hand, analytical expressions require cumbersome work due to complexity of these coupled equations. For the purpose of this work, we approximate the STM. Basically, when $t-t_0$ is very small, the state transition matrix can be approximated by series expansion as shown below
\begin{multline}
\boldsymbol{\Phi}(t_k,t_0)=\begin{bmatrix}
I & 0 \\ 
0 & I
\end{bmatrix} + \begin{bmatrix}
0 & I \\ 
G & 2\Omega
\end{bmatrix} \Delta t + \begin{bmatrix}
G & 2\Omega \\ 
2\Omega G & G + 4\Omega^2
\end{bmatrix} \frac{\Delta t^2}{2} + \begin{bmatrix}
2 \Omega G & G + 4\Omega^2 \\ 
G^2 + 4\Omega^2 G & 2\Omega G+ 2 G \Omega + 8\Omega^2
\end{bmatrix} \frac{\Delta t^3}{6} \\ 
+ \begin{bmatrix}
G^2 + 4\Omega^2 G & 2\Omega G + 4\Omega G \Omega + 4 G \Omega^2 + 16\Omega^3 \\ 
2\Omega G^2+ 2G\Omega G+8 \Omega^2 G & G^2 + 4\Omega^2 G + 4\Omega G \Omega + 4G\Omega^2 + 16\Omega^3
\end{bmatrix} \frac{\Delta t^4}{24} + \mathcal{O}(A^5)
\end{multline}

where $\Delta t = t_k-t_0$. The Taylor expansion for the $3 \times 3$ position and velocity sub-matrices of the \ac{STM} by ignoring the higher order terms are:
\begin{eqnarray}
    \Phi_{rr} = \frac{\partial \mathbf{r}}{\partial \mathbf{r}_0} = I + G \frac{\Delta t^2}{2} + (2\Omega G)\frac{\Delta t^3}{6} + (G^2+4\Omega^2G)\frac{\Delta t^4}{24}
\end{eqnarray}
\begin{eqnarray}
\label{eqnphirv}
    \Phi_{rv}= \frac{\partial \mathbf{r}}{\partial \mathbf{v}_0} =  I \Delta t + \Omega \Delta t^2 + (G + 4 \Omega ^2)\frac{\Delta t^3}{6} + (2\Omega G + 4\Omega G \Omega + 4G\Omega^2 + 16\Omega^3)\frac{\Delta t^4}{24}
\end{eqnarray}
\begin{eqnarray}
       \Phi_{vr} = \frac{\partial \mathbf{v}}{\partial \mathbf{r}_0} =  I+ G \Delta t + \Omega G \Delta t^2 + (G^2 +4\Omega^2 G) \frac{\Delta t^3}{6} + (2\Omega G^2+ 2G  \Omega G + 8\Omega^2 G)\frac{\Delta t^4}{24}
\end{eqnarray}
\begin{eqnarray}
    \Phi_{vv}\hspace{-0.5mm}=\hspace{-0.5mm}\frac{\partial \mathbf{v}}{\partial \mathbf{v}_0} \hspace{-0.5mm}=\hspace{-0.5mm}2\Omega \Delta t + (G + 4\Omega ^2) \frac{\Delta t^2}{2} + (2 \Omega G + 2 G\Omega\hspace{-0.5mm}+\hspace{-0.5mm}8\Omega^2) \frac{\Delta t^3}{6}\hspace{-0.5mm}+\hspace{-0.5mm}(G^2 +4\Omega^2 G + 4 \Omega G \Omega + 4 G\Omega^2 +16\Omega^3)\frac{\Delta t^4}{24}
\end{eqnarray}

Being G the gradient matrix at the end of the propagation interval. Finally, an approximation is performed by ignoring the remaining terms: 
\begin{eqnarray}
\label{eqnPhi}
\Phi(t_k,t_0) \approx  \begin{bmatrix}
\Phi_{rr} & \Phi_{rv} \\ 
\Phi_{vr} & \Phi_{vv}
\end{bmatrix}
\end{eqnarray}

A performances comparison of Eq.\ref{eqnPhi} with respect to the numerically derived \ac{STM} is given in~\ref{appendixb}. It should be noted that Eq.\ref{eqnPhi} is given for only one of the spacecraft, requiring calculations for both spacecraft, $i=1,2$:
\begin{eqnarray}
\Phi(t_k,t_0)=\begin{bmatrix}
\Phi_1(t_k,t_0) & 0_{6 \times 6} \\ 
0_{6 \times 6} & \Phi_2(t_k,t_0)
\end{bmatrix}
\end{eqnarray}

$\boldsymbol{\tilde H}$ can be written in the following form of derivation with respect to the position and velocity vectors considering crosslink range observations:
\begin{eqnarray}
\boldsymbol{\tilde H}=\begin{bmatrix}
\frac{\partial \mathbf{\rho}}{\partial \mathbf{r_1}} & \frac{\partial \mathbf{\rho}}{\partial \mathbf{v_1}} & \frac{\partial \mathbf{\rho}}{\partial \mathbf{r_2}} & \frac{\partial \mathbf{\rho}}{\partial \mathbf{v_2}}
\end{bmatrix}=\begin{bmatrix}
\frac{\partial \mathbf{\rho}}{\partial \mathbf{r_1}} & 0_{1 \times 3} & \frac{\partial \mathbf{\rho}}{\partial \mathbf{r_2}} & 0_{1 \times 3}
\end{bmatrix}
\end{eqnarray}

Now, we can directly calculate Eq. (\ref{Hmtx}), $\boldsymbol{{H}}_k = \boldsymbol{\tilde{H}}_k \boldsymbol{\Phi}(t_k,t_0)$:
\begin{eqnarray}
\label{eqnHcompact}
    \boldsymbol{{H}} = \boldsymbol{\tilde{H}} \boldsymbol{\Phi}(t_k,t_0) = \begin{bmatrix}
\frac{\partial \mathbf{\rho}}{\partial \mathbf{r_1}} & 0_{1 \times 3} & \frac{\partial \mathbf{\rho}}{\partial \mathbf{r_2}} & 0_{1 \times 3}
\end{bmatrix}\begin{bmatrix}
\begin{bmatrix}
\Phi_{1,rr} & \Phi_{1,rv} \\ 
\Phi_{1,vr} & \Phi_{1,vv}
\end{bmatrix} & 0_{6 \times 6}\\ 
0_{6 \times 6} & \begin{bmatrix}
\Phi_{2, rr} & \Phi_{2,rv} \\ 
\Phi_{2,vr} & \Phi_{2,vv}
\end{bmatrix}
\end{bmatrix}\\=
\begin{bmatrix}
\frac{\partial \mathbf{\rho}}{\partial \mathbf{r_1}}\Phi_{1, rr} & \frac{\partial \mathbf{\rho}}{\partial \mathbf{r_1}}\Phi_{1, rv} & \frac{\partial \mathbf{\rho}}{\partial \mathbf{r_2}}\Phi_{2, rr} & \frac{\partial \mathbf{\rho}}{\partial \mathbf{r_2}}\Phi_{2, rv}
\end{bmatrix}
\end{eqnarray}

In Eq.\ref{eqnHcompact}, the velocity related terms for both spacecraft are $\frac{\partial \mathbf{\rho}}{\partial \mathbf{r_1}}\Phi_{1, rv}$ and $\frac{\partial \mathbf{\rho}}{\partial \mathbf{r_2}}\Phi_{2, rv}$ respectively. As it can be seen from Eq.\ref{eqnphirv}, there is a null gradient matrix $\boldsymbol{G}(t)$ related terms in $\Phi_{rv}$ till the third order expansion (e.g. $\Phi_{rv}=I\Delta t + \Omega \Delta^2$). This is why velocity related terms in $\boldsymbol{H}$ are linearly dependent at the second order approximation. On the other hand, many columns of $\boldsymbol{\tilde H}_{k}$ are equal in magnitude and opposite in sign, $\tilde H_k^1 = -\tilde H_k^7$ ($\frac{\partial\dot{\rho}}{\partial x_1}=-\frac{\partial\dot{\rho}}{\partial x_2}$). This would tend to make the rows of the information and observability matrices dependent \cite{hillthesis}. As stated in \cite{hillthesis}, the differences in the state transition matrix, $\boldsymbol{\Phi}(t_k,t_0)$, when one of the spacecraft has a unique orbit, make $\boldsymbol{{H}}_k$ and thus the information  and observability matrices positive definite. These differences can be seen, for example, in $H_k^1$ as partial derivatives of ${\ddot{x}}_1,{\ddot{y}}_1,{\ddot{z}}_1$ with respect to $x_1$ and in $H_k^7$ as partial derivatives of ${\ddot{x}}_2,{\ddot{y}}_2,{\ddot{z}}_2$ with respect to $x_2$. During the orbital trajectory, \ac{STM} will be unique in cislunar vicinity and this will bring an observable system.

Now both Eq.\ref{Neqn}, $\boldsymbol{N}$ and Eq.\ref{infomatrix}, $\boldsymbol{\Lambda}$, can be built. By remembering that the inverse of the information matrix is nothing but the covariance matrix, $\boldsymbol{P}=\boldsymbol{\Lambda}^{-1}$. \textit{A priori} covariance, $\boldsymbol{P}_0$, may artificially make the system observable so that $\boldsymbol{P}_0$ should not be added in the observability analysis. Lastly, the information matrix at $t_k$ alone, $\boldsymbol{H}^T \boldsymbol{W} \boldsymbol{H}$, in other words the effectiveness of observation at $t_k$, can be derived by using Eq.\ref{eqnHcompact}.  For the observation effectiveness of the position states alone, $x_i, y_i, z_i$ of the spacecraft $i=1,2$, can be calculated as:
\begin{eqnarray}
    \delta \Lambda_{\mathbf{r}, ij}(t_k)= \frac{1}{\sigma_{\rho}} \sqrt{\text{row}_j(\Phi^T_{i, rr} (t_k, t_0))(\frac{\partial \rho}{\partial \mathbf{r}_i})^T (\frac{\partial \rho}{\partial \mathbf{r}_i}) \text{col}_j (\Phi_{i, rr}(t_k, t_0))}
\end{eqnarray}

and for the velocity states alone, $x_i, y_i, z_i$ of the spacecraft $i=1,2$:
\begin{eqnarray}
    \delta \Lambda_{\mathbf{v}, ij} (t_k)= \frac{1}{\sigma_{\rho}} \sqrt{\text{row}_j(\Phi^T_{i, rv}(t_k, t_0))(\frac{\partial \rho}{\partial \mathbf{r}_i})^T (\frac{\partial \rho}{\partial \mathbf{r}_i}) \text{col}_j (\Phi_{i, rv}(t_k, t_0))}
\end{eqnarray}

where subscript $j$ indicates element number (e.g. $j=1$ for $x$ as a first state). As it can be seen, the observation effectiveness, thus the estimation accuracy, is directly related to the direction of observation and certain rows of the \ac{STM} which is related to the divergent dynamics along a trajectory. Basically, \ac{STM} approximates how a slight deviation in state variables propagates along the trajectory and it would be better from the estimation perspective if measurement vectors are not perpendicular to these deviations.

In this subsection $\boldsymbol{{\tilde H}}_k$ is given for range, range-rate, and \ac{LOS} angle measurements. However, $\boldsymbol{{H}}_k$ is shown for the range measurement only case due to complexity of other measurement models. The first element of $\boldsymbol{{H}}_k^{\dot \rho_{1}}$ is given below as an example. In a similar way, other Jacobian matrices, $\boldsymbol{{H}}_k^{\dot \rho}$, $\boldsymbol{{H}}_k^{\phi}$ and $\boldsymbol{{H}}_k^{\varphi}$, can be calculated.
\begin{multline}
    \boldsymbol{{H}}_k^{\dot \rho_{1}}=\left(\frac{\partial{\ddot{x}}_k}{\partial x_k}\frac{{T_s}^2}{2}+1\right)\left(\frac{\left({\dot{x}}_1-{\dot{x}}_2\right)\rho^2-\left(x_1-x_2\right)\ \boldsymbol{\rho}\cdot\dot{\boldsymbol{\rho}}}{\rho^3}\right)+\left(\frac{\partial{\ddot{y}}_k}{\partial x_k}\frac{{T_s}^2}{2}\right)\left(\frac{\left({\dot{y}}_1-{\dot{y}}_2\right)\rho^2-\left(y_1-y_2\right)\ \boldsymbol{\rho}\cdot\dot{\boldsymbol{\rho}}}{\rho^3}\right)+ \\ \left(\frac{\partial{\ddot{z}}_k}{\partial x_k}\frac{{T_s}^2}{2}\right)\left(\frac{\left({\dot{z}}_1-{\dot{z}}_2\right)\rho^2-\left(z_1-z_2\right)\ \boldsymbol{\rho}\cdot\dot{\boldsymbol{\rho}}}{\rho^3}\right)+\left(\frac{\partial{\ddot{y}}_k}{\partial x_k}{T_s}^2+\frac{\partial{\ddot{x}}_k}{\partial x_k}T_s\right)\left(\frac{\left(x_1-x_2\right)}{\rho}\right)+\\\left(\frac{\partial{\ddot{x}}_k}{\partial x_k}{T_s}^2+\frac{\partial{\ddot{y}}_k}{\partial x_k}T_s\right)\left(\frac{\left(y_1-y_2\right)}{\rho}\right)+ \left(T_s\frac{\partial{\ddot{z}}_k}{\partial x_k}\right)\left(\frac{\left(z_1-z_2\right)}{\rho}\right)
\end{multline}

\subsection{Consider Covariance Analysis}

For crosslink based autonomous navigation systems, it is expected a certain level of bias in the radiometric measurements, predominantly because of the transmit and receive line delays on both spacecraft. Bias errors in the estimation process can be investigated in different ways. This can either be neglected or estimated by including dynamic or measurement model parameters into the state vector. Also, it can be considered by assuming bias is constant and its \textit{a priori} estimate and associated covariance matrix are known. The consider approach is investigated in detail by many researchers (\cite{Lou2015, montenbruck2002satellite, Grewal2014}). The consider covariance or filter analysis can be in a batch or in a sequential form. In this study, the sequential form of the consider filter is studied considering the specific case of time invariant measurement bias. This means the consider parameter $c_k = c_{k-1}$ for all $k$, and the bias covariance matrix $P_k^{cc}$ is time invariant. The sequential consider covariance filter or the \ac{SKF} equations can be found in \cite{Lou2015}.

\subsection{Monte Carlo Analysis}

Monte Carlo methods are commonly used for sensitivity analysis and quantitative probabilistic analysis in the navigation system design especially for analyzing errors. In this study, the \ac{RMS} error, in each $k$th time step, for the $N$th case of the Monte Carlo simulation is calculated by using following equation:
\begin{eqnarray}
\label{eqnmontecarlo}
    {RMSE}_{k}=\sqrt{\frac{1}{N}\sum_{i=1}^{N}(x_{i,k}-\hat{x}_{i,k})^{2}}
\end{eqnarray}
where $x_{i,k}$ and $\hat{x}_{i,k}$ are $i$th component of state vector and its estimate respectively.

\subsection{Observation Type Comparison}

This study investigates the radiometric observables which are range, range-rate, and \ac{LOS} angles. In order to test a navigation system, it should be studied which one of the observables provides better navigation performances for the same radio measurement system. In this section, the relation between radiometric observables will be given from the measurement precision point of view. In section~\ref{LOSsection}, it is already shown how \ac{LOS} angles could be derived from range measurements. This is why in this section the relation between range and range-rate measurements is given only. A quantitative approach is given in \cite{Dirkx2018} and a similar method has been used in this section. A \ac{SNR} criterion for an observable $h$ ($\rho$ or $\dot \rho$) and estimated states $\mathbf{X}$ can be given as:
\begin{eqnarray}
    SNR_{h,k}=\left | \frac{1}{\sigma_h} \boldsymbol{H}_k \right |
\end{eqnarray}
where $\sigma_h$ is the noise level of the measurement $h$. Basically, we define the following figure of merit to compare the relative sensitivity of range and range-rate observables to estimated states $\mathbf{X}$:
\begin{eqnarray}
\label{eqnfom}
    \Xi_\textbf{X}=\frac{SNR_\rho}{SNR_{\dot{\rho}}}
\end{eqnarray}
If $\Xi_{\textbf{X}}<1$, it can be said range-rate observation would become a feasible alternative to range observation for estimating $\textbf{X}$. This can be applied to other states as well. The method given here can be considered as a ratio of the observability Gramian at each time epoch, and multiplying it by a realistic relative measurement error parameter. Before making a realistic comparison, we would need to find the relation between range and range-rate observation errors, in other words relative error parameter. At first, the ranging error will be defined. In case a conventional tone ranging is assumed for ranging operations, the following can be used to calculate the phase error on the major tone in radians:
\begin{eqnarray}
    \sigma_r=\sqrt{\frac{2B_n}{2\frac{S}{N_0}}}
\end{eqnarray}
where $2B_n$ Bi-lateral loop bandwidth, $S/N_0$, signal-to-noise ratio in dB Hz. Basically, the phase noise on the major ranging tone results in a range measurement error by multiplying the wavelength of the ranging signal, $\lambda/2\pi$ as:
\begin{eqnarray}
\label{eqnsigmarange}
    \sigma_\rho = \sigma_r \frac{\lambda}{2\pi}
\end{eqnarray}
As an example, a \SI{30}{dB Hz} signal-to-noise ratio on the major tone with \SI{0.1}{Hz} loop bandwidth results in a \SI{0.32}{m} ranging error.

Doppler data noise can be expressed by the phase noise in radians and converted to range-rate noise by following equation \cite{montenbruck2002satellite}:
\begin{eqnarray}
\label{eqnsigmarangerate}
    \sigma_{\dot{\rho}}=\frac{\sqrt2c}{2\ G\ f_t\ t_c}\frac{\sigma_\varphi}{2\pi}
\end{eqnarray}
with $c$ speed of light, $G$ transponding ratio, $f_t$, transmitted frequency, $t_c$, integration time, $\sigma_\varphi$ phase noise in radians. Thus, the ratio of range and range-rate error, $\zeta$, can be found by dividing Eq.~\ref{eqnsigmarange} to~\ref{eqnsigmarangerate}:
\begin{eqnarray}
\label{eqnzeta}
    \zeta=\frac{\sigma_\rho}{\sigma_{\dot{\rho}}}=\sqrt2\ G\ \frac{f_t}{f_{MT}}t_c
\end{eqnarray}
As an example, for an S-band system, \SI{1}{m} ranging error would be equal to \SI{0.3}{mm/s} range-rate error with \SI{1}{s} integration on the same ranging/Doppler unit.

By using Eq.~\ref{eqnzeta}, it is possible to compare range only and range-rate only navigation systems in a realistic way. Now, Eq.~\ref{eqnfom} can be rewritten as:
\begin{eqnarray}
\label{xieqn}
    \Xi_k^i=\frac{1}{\zeta}\left|\frac{H_\rho^i}{H_{\dot{\rho}}^i}\right|
\end{eqnarray}
where $i$ represents state number and $k$ represents time. 
\section{Results}
\label{sec6}
\begin{figure}[h]
    \centering
    \includegraphics[width=0.33\textheight]{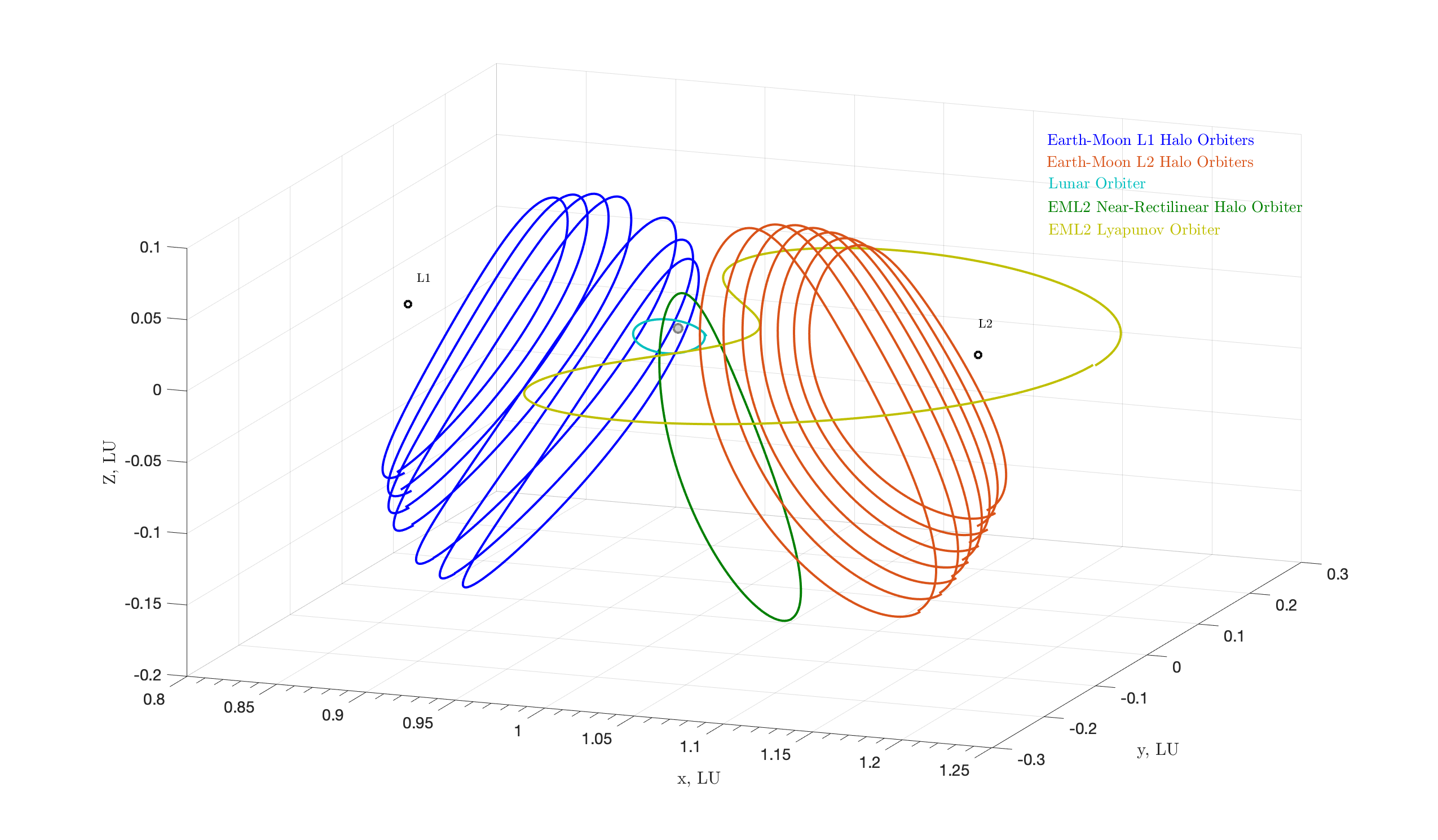}
    \caption{Orbital trajectories considered in this study}
    \label{fig:traj}
\end{figure}

This section presents the effects of the various parameters on the performances of autonomous navigation system. Based on various orbital trajectories, shown in Figure~\ref{fig:traj}, measurement type, accuracy, bias, frequency, formation geometry and network topology have been investigated.

\subsection{Measurement type}

\begin{figure}[h]
    \centering
    \includegraphics[width=0.66\textheight]{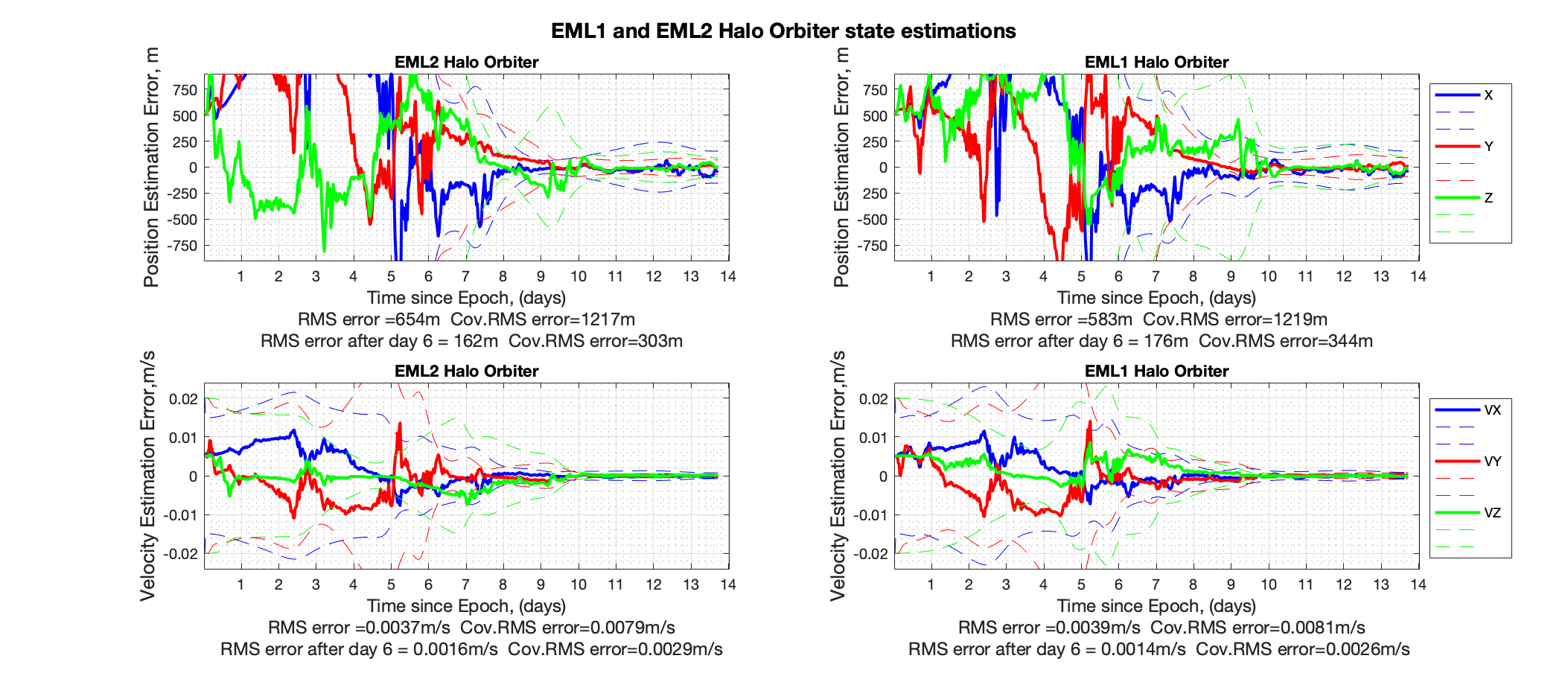}
    \caption{EML1 and EML2 Halo Orbiter state estimation based on \SI{1}{m} ranging error}
    \label{fig:4lu}
\end{figure}

Autonomous navigation method works well with Lagrangian point orbiters. In order to show that a link between L1 and L2 Halo orbiters has been set and based on \SI{1}{m} ranging error, 14 days of simulation has been executed. Corresponding estimation results can be seen in Figure~\ref{fig:4lu}. Estimation errors are in order of \SI{100}{m} for position and \SI{1}{mm/s} for velocity. This shows even in less observable geometrical conditions, the navigation method works well. 

\begin{figure}[h]
    \centering
    \includegraphics[width=0.33\textheight]{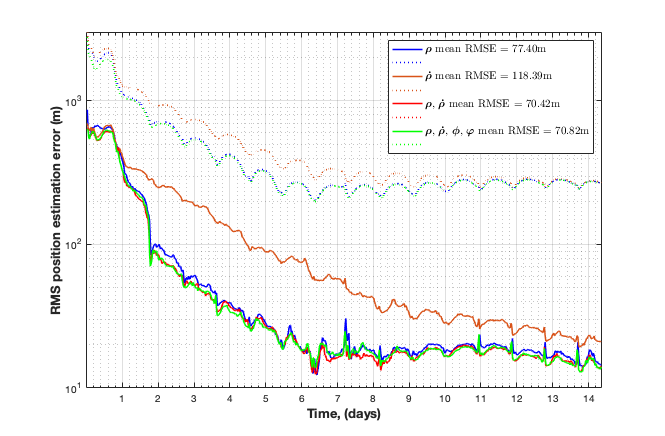}
    \caption{\ac{RMS} position estimation error for four different measurement type scenario. The dashed lines represents $3\sigma$ covariance bounds}
    \label{fig:posmeanrms}
\end{figure}

\begin{figure}[h]
    \centering
    \includegraphics[width=0.33\textheight]{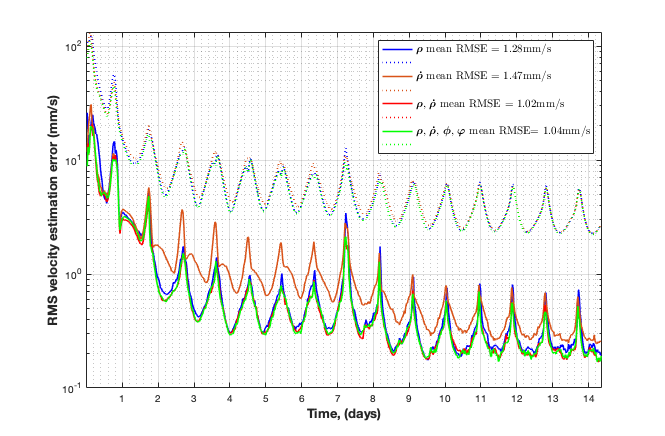}
    \caption{\ac{RMS} velocity estimation error for four different measurement type scenario}
    \label{fig:velmeanrms}
\end{figure}

The purpose of this section is to illustrate the effect of the different measurement types (range, range-rate, and LOS angle) on the estimation accuracy. First, all three measurement types have been compared in the same scenario: one of the spacecraft orbits around the Moon and the second one with an Earth-Moon L2 Southern Halo orbit. Measurement accuracy has been adjusted in a realistic way by using Eq.\ref{eqnzeta} for range and range-rate, and Eq.\ref{eqntimedelay} for LOS angles: the errors considered respectively are \SI{1}{m}, \SI{0.3}{mm/s}, and \SI{0.5}{deg}. No bias has been assumed for measurement types. Figure~\ref{fig:posmeanrms} shows the average position estimation error (including both spacecraft position states) for a Monte Carlo simulation (100 executions) including $3\sigma$ covariance bounds. The velocity estimation can be seen in Figure~\ref{fig:velmeanrms}. The average position and velocity errors have been estimated by taking the mean value of the corresponding spacecraft states after running the \ac{RMS} errors at the each time step by using Eq.\ref{eqnmontecarlo}. As it can be seen, using range and range-rate observations provides the best state estimation results. On the other hand, having \ac{LOS} angle measurements didn't improve the performances at all. As these measurements contains high errors and range measurements provide more useful data to the filter. Feeding the filter with very accurate \ac{LOS} measurements slightly improves the navigation system performances, though. In the case a \SI{0.001}{deg} measurements error is considered (instead of \SI{0.5}{deg}), the \ac{RMS} position and velocity estimation error would be \SI{57.54}{m} and \SI{0.92}{mm/s} respectively. However, this would require very a accurate measurement system. This scenario illustrates what would happen in case other types of accurate \ac{LOS} sensors had been considered. On the other hand, combining range and range-rate measurements improved the estimation performance by almost $10\%$ as compared to range only. This means that for certain geometries, range-rate measurements provide useful information to the navigation filter. This trend has been observed in the velocity estimation as well. In addition, a EML1-L2 scenario has been investigated and all results can be seen in Table~\ref{tab:meastype}. 

\begin{table}
\label{tab:meastype}
\centering
\caption{RMS State estimation errors based on 100 Monte-Carlo executions}
\begin{tabular}{|l|l|c|c|} 
\hline
\multicolumn{2}{|l|}{} & EML1-L2 Orbiters & EML2-Lunar Orbiters \\ 
\hline
\multirow{2}{*}{Range-only ($\rho$)} & RMS Pos.Err. (m) & 487.65 & 77.40 \\ 
\cline{2-4}
 & RMS Vel.Err. (mm/s) & 2.85 & 1.28 \\ 
\hline
\multirow{2}{*}{Range-rate-only ($\dot{\rho}$)} & RMS Pos.Err. (m) & 803.63 & 118.39 \\ 
\cline{2-4}
 & RMS Vel.Err. (mm/s) & 4.66 & 1.47 \\ 
\hline
\multirow{2}{*}{Range and range-rate ($\rho, \dot{\rho}$)} & RMS Pos.Err. (m) & 483.68 & 70.42 \\ 
\cline{2-4}
 & RMS Vel.Err. (mm/s) & 2.82 & 1.02 \\ 
\hline
\multirow{2}{*}{Range, range-rate and LOS ($\rho, \dot{\rho}, \phi, \varphi$)} & RMS Pos.Err. (m) & 486.14 & 70.82 \\ 
\cline{2-4}
 & RMS Vel.Err. (mm/s) & 2.85 & 1.04 \\
\hline
\end{tabular}
\end{table}

\begin{figure}[h]
    \centering
    \includegraphics[width=0.33\textheight]{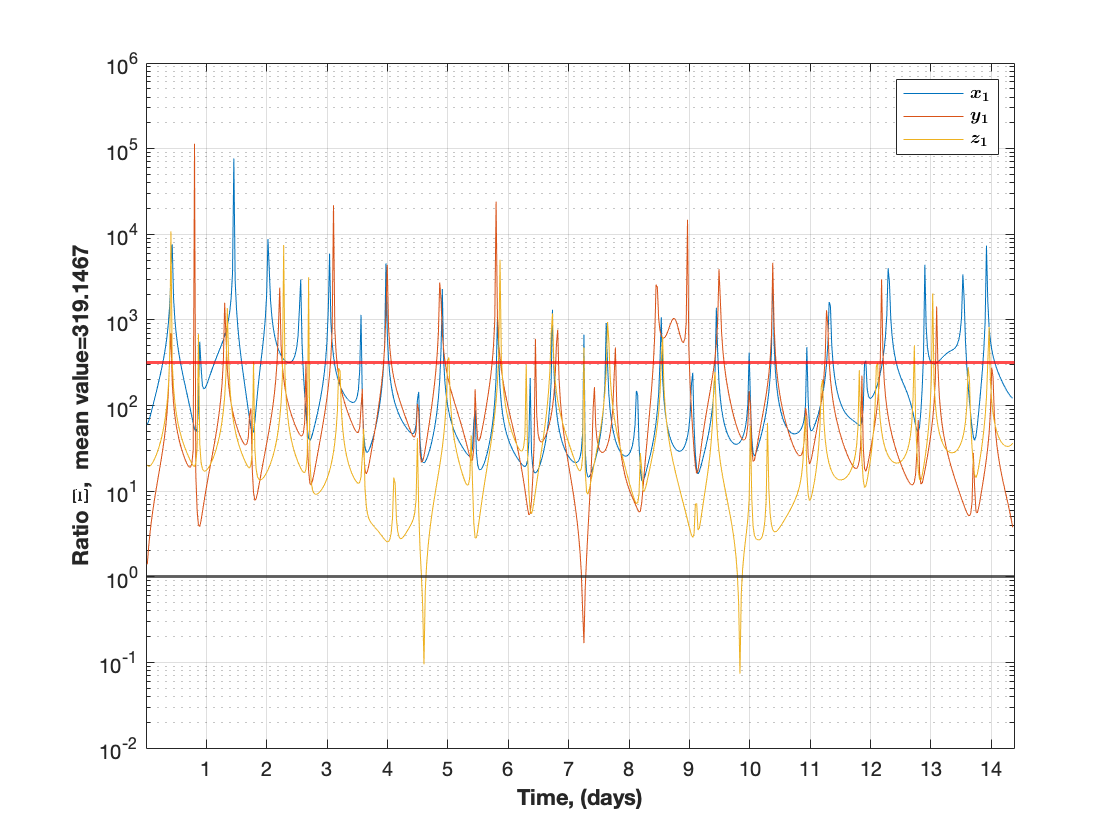}
    \caption{Ratio $\Xi$ values for position states of the Earth-Moon L2 Halo Orbiter. Horizontal black and red lines represent the threshold value and the mean value, respectively.}
    \label{fig:xipossc1}
\end{figure}

\begin{figure}[h]
    \centering
    \includegraphics[width=0.33\textheight]{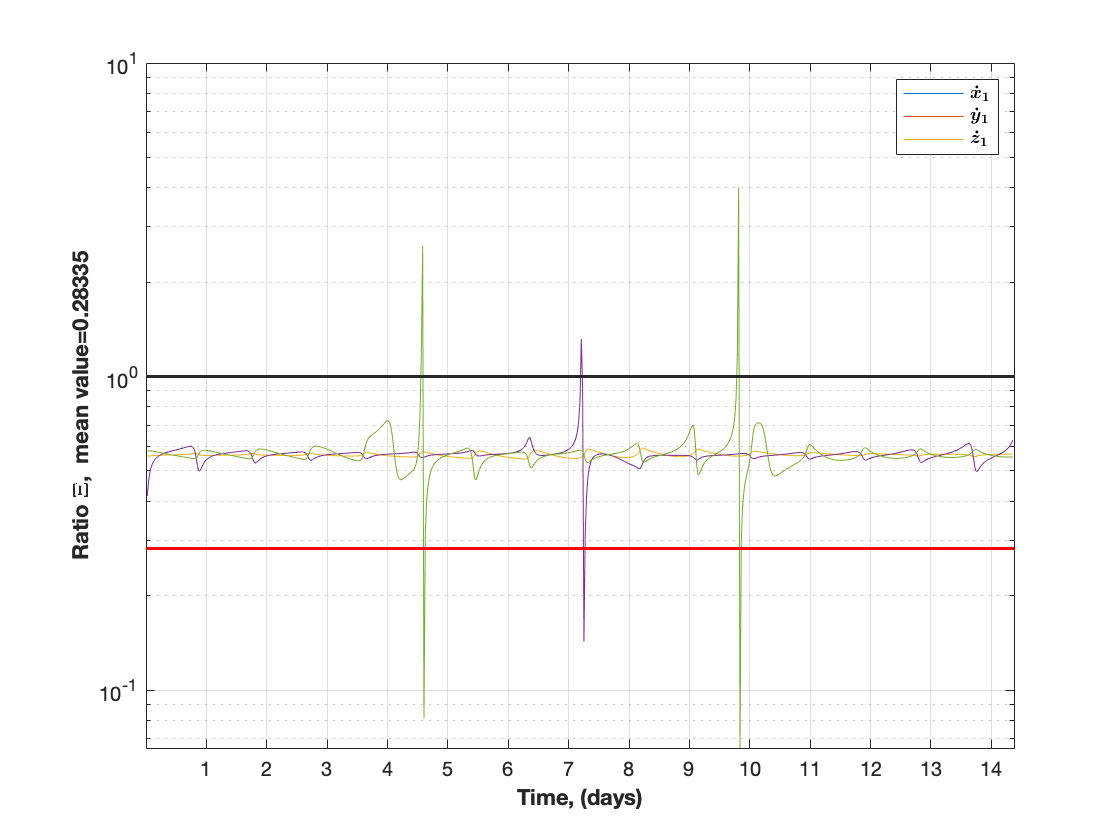}
    \caption{Ratio $\Xi$ values for velocity states of the Lunar Orbiter. Horizontal black and red lines represent the threshold value and the mean value, respectively.}
    \label{fig:xivelsc2}
\end{figure}

For the same scenario, the ratio $\Xi$ has been plotted for the position states of the Earth-Moon L2 Halo orbiter, $x_1, y_1, z_1$, in Figure~\ref{fig:xipossc1} with a horizontal red line representing the overall mean value of them. As it can be seen, almost all the times, the ratio $\Xi$ is higher than the threshold, ($\Xi > 1$). The average value of $\Xi$ for this case can also been seen as $319.14$, which means the range-only case with a measurement error of \SI{1}{m} provides better position estimation for the EML2 orbiter than the range-rate-only case with a measurement error of \SI{0.3}{mm/s}. On the other hand, the range-rate-only case provides better velocity estimation with a mean ratio $\Xi$ of $0.28$ (See Figure~\ref{fig:xivelsc2}). Considering other states, in brief, the range only case shows better performances than range-rate only case.

In the second case, only range and range-rate measurements have been compared for different orbital configurations. In each different orbital scenario, four different cases have been investigated for a full orbital period. Based on \SI{1}{m} and \SI{0.3}{mm/s} measurement errors, $\Xi_{mean}$ values are computed. All the mean results, $\Xi_{mean}$, are given in Table~\ref{xitable}. All the results are higher than the threshold, $\Xi > 1$, which means range only cases provide better overall state estimation than range-rate only cases. A configuration in which spacecraft are located at  EM L1 and L2 showed slightly higher $\Xi_{mean}$ results than the other cases. This means that range-rate measurements would provide relatively less information to the filter than range measurements in case the formation is made by spacecraft at both Lagrangian points. However, in case one of the spacecraft orbits around the Moon, range-rate measurements would provide relatively higher information than the previous case. However, this still isn't sufficient for the range-rate-only case to have the same performances with respect to the range-only case. 

Overall, range measurements would be more beneficial than range-rate measurements for  radiometric autonomous navigation of cislunar small satellite formations. On the other hand, combining almost perfect \ac{LOS} angles with range measurements did not improve the navigation performances. This shows that crosslink range measurements are much better performing than angles as already noted in~\cite{hillthesis}. 

\begin{table}[h]
\centering
\caption{Measurement type comparison for different orbital configurations}
\label{xitable}
\resizebox{\columnwidth}{0.65cm}{%
\begin{tabular}{|c|c|c|c|c|c|c|c|c|c|c|c|c|} 
\hline
\multirow{2}{*}{} & \multicolumn{4}{c|}{EML1/L2} & \multicolumn{4}{c|}{EML1/Lunar} & \multicolumn{4}{c|}{EML2/Lunar} \\ 
\cline{1-13}
 Halo Period (TU) & $T_{2.786/3.306}$ & $T_{2.778/3.276}$ & $T_{2.759/3.240}$ & $T_{2.721/3.195}$ & $T_{2.786}$ & $T_{2.778}$ & $T_{2.759}$ & $T_{2.721}$ & $T_{3.306}$ & $T_{3.276}$ & $T_{3.240}$ & $T_{3.195}$ \\ 
\hline
$\Xi_{mean (\SI{1}{m}/\SI{0.3}{mm/s})}$ & $289.21$ & $365.53$ & $510.73$ & $219.14$ & $182.65$ & $155.05$ & $221.02$ & $75.31$ & $110.86$ & $116.50$ & $167.17$ & $116.25$ \\
\hline
\end{tabular}%
}
\end{table}

\subsection{Measurement accuracy}

Measurement errors also affect the orbit determination accuracy. It is expected that higher measurement errors result in higher orbit determination errors. In order to see how the orbit determination accuracy changes with the increasing measurement errors, Monte Carlo simulations with 100 executions have been performed for various orbital configurations. In these simulations, range only and range-rate only cases have been investigated. Based on the link between two spacecraft at three different orbits, Earth-Moon L2 Southern Halo, L1 Southern Halo, Lunar, three different measurement accuracies for each measurement type (range, range-rate), \SI{1}{m}, \SI{10}{m}, \SI{100}{m} and \SI{0.1}{mm/s}, \SI{1}{mm/s}, \SI{10}{mm/s}, respectively, have been studied in this section. 

\begin{figure}[h]
    \centering
    \includegraphics[width=0.33\textheight]{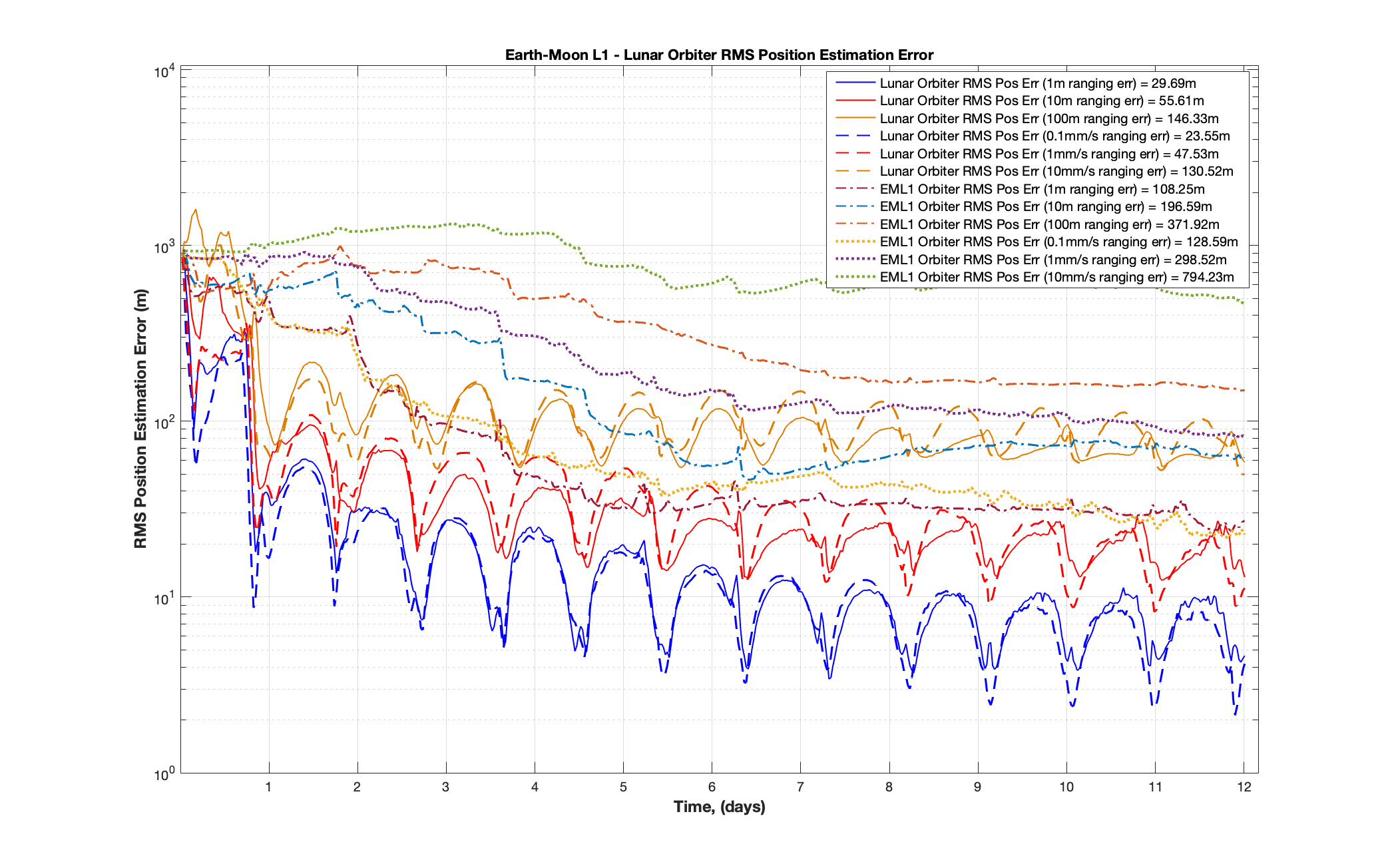}
    \caption{The Earth Moon L1 - Lunar Orbiter case RMS position estimation error for various range and range-rate measurement errors}
    \label{fig:accPos1}
\end{figure}

\begin{figure}[h]
    \centering
    \includegraphics[width=0.33\textheight]{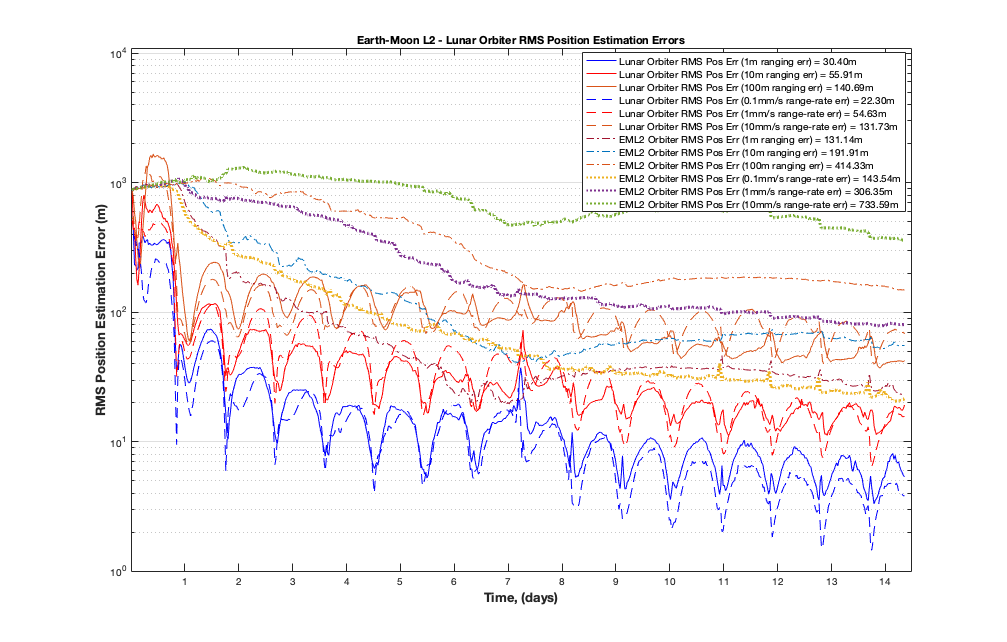}
    \caption{The Earth Moon L2 - Lunar Orbiter case RMS position estimation error for various range and range-rate measurement errors}
    \label{fig:accPos2}
\end{figure}

\begin{figure}[h]
    \centering
    \includegraphics[width=0.33\textheight]{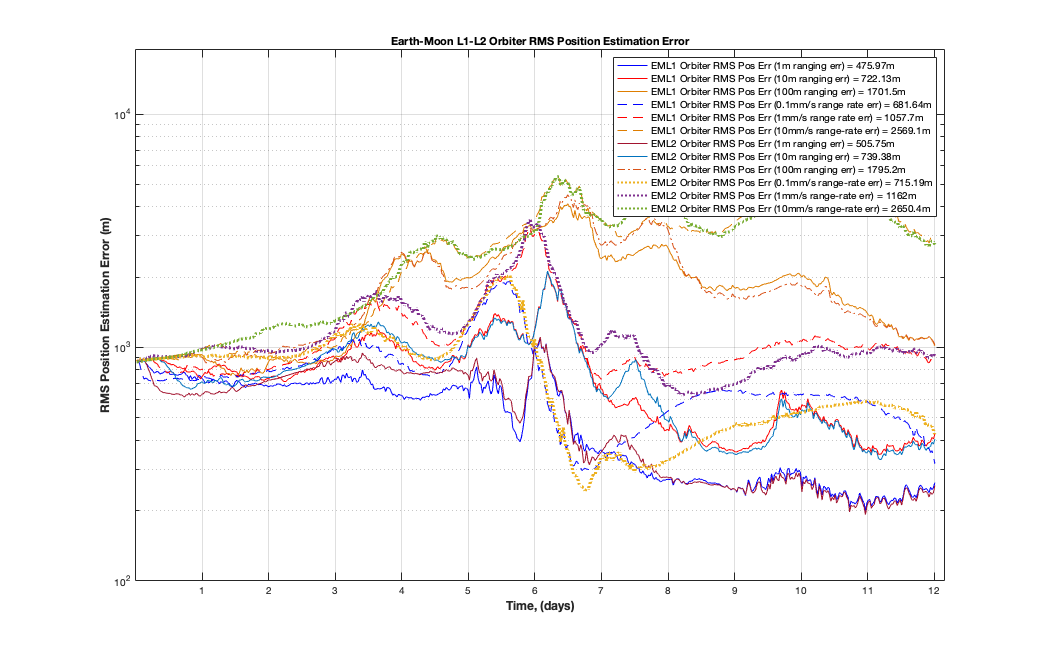}
    \caption{The Earth Moon L1 - L2 Orbiter case RMS position estimation error for various range and range-rate measurement errors}
    \label{fig:accPos3}
\end{figure}

The Monte Carlo based \ac{RMS} position estimation results can be seen Figures~\ref{fig:accPos1},~\ref{fig:accPos2} and~\ref{fig:accPos3} for Earth-Moon L1-Lunar, Earth-Moon L2-Lunar, and Earth-Moon L1-L2 cases respectively. As it can be seen, higher measurement errors result in less accurate state estimation. Lunar and Lagrangian orbiter cases have almost the same order of magnitude position estimation errors. On the other hand, the EM L1-L2 case has higher estimation errors when the measurement errors are increasing. This is due to the fact that the system is less observable and the condition number and the unobservability index for both range and range-rate observations please see the Table~\ref{measacctable} for further details) are higher for this case than the others. Basically, less observable states are affected more in case of high measurement errors. Overall, high observable systems, like the EM L1-Lunar and EML2-Lunar Orbiter cases, have been less affected by high measurement errors. This means that less accurate inter-satellite measurement methods, such as time-derived or data-aided ranging, could be an option for these type of mission configurations considering the typical design challenges in small satellites \cite{Turan2022}. 

\begin{table}
\centering
\caption{Condition number and unobservability index for various measurement types and corresponding cases}
\label{measacctable}
\begin{tabular}{|c|c|c|c|c|} 
\hline
 &  & EML1-Lunar & EML2-Lunar & EML1-L2 \\ 
\hline
\multirow{2}{*}{Cond.num} & Range & \SI{1.34e10}{}  & \SI{1.10e10}{}  & \SI{3.57e10}{}  \\ 
\cline{2-5}
 & Range-rate & \SI{2.35e10}{}  & \SI{2.53e10}{}  & \SI{3.82e10}{}  \\ 
\hline
\multirow{2}{*}{Unobs. index} & Range & \SI{3.82e2}{}  & \SI{2.00e2}{}  & \SI{1.81e4}{}  \\ 
\cline{2-5}
 & Range-rate & \SI{0.2134}{}  & \SI{0.1756}{}  & \SI{2.84e3}{}  \\
\hline
\end{tabular}
\end{table}

\subsection{Measurement bias}

\begin{figure}[h]
    \centering
    \includegraphics[width=0.5\textheight]{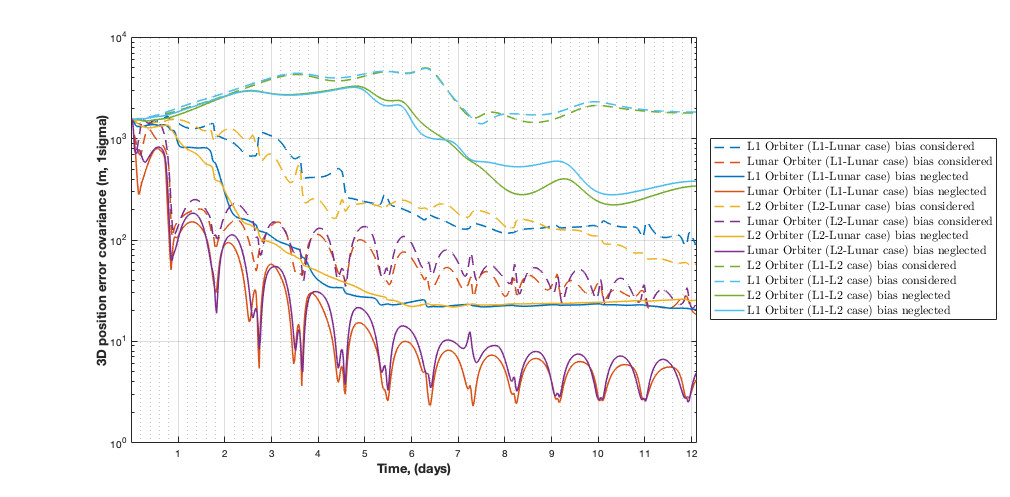}
    \caption{Position uncertainty (1$\sigma$) for three-different cases with \SI{20}{m} ranging bias considered, compared with neglecting bias.}
    \label{fig:biasconsidered}
\end{figure}

In this section the consider covariance results achieved are presented. The effect of considering a \SI{20}{m} ranging bias for three different mission scenarios: EM L1-L2 orbiter, EM L1-Lunar orbiter and EM L2-Lunar orbiter can be seen in Figure~\ref{fig:biasconsidered}. 

In the considered bias scenarios, the \ac{RMS} error covariance increased up to $85 \%$ for the case of Lagrangian orbiters and $125 \%$ for the cases of Lagrangian - Lunar orbiters. 

\subsection{Measurement frequency}

\begin{figure}[h]
    \centering
    \includegraphics[width=0.33\textheight]{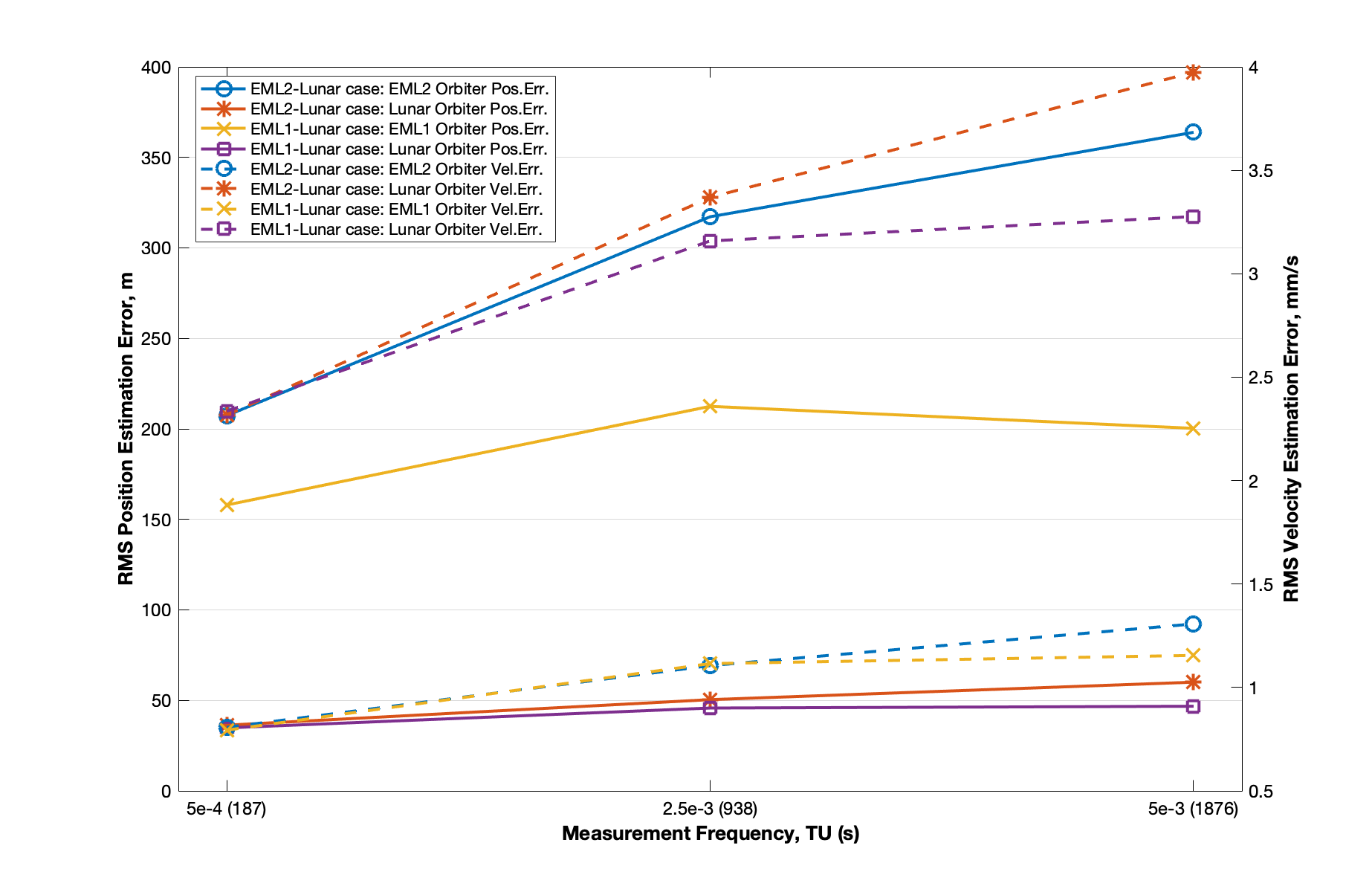}
    \caption{The Earth Moon L1 - L2 Orbiter case RMS position/velocity estimation errors for various measurement frequencies}
    \label{fig:measfreq}
\end{figure}

It is expected that increasing the measurement interval would result in increased time of filter convergence and increased \ac{RMS} error with respect to the case of the same amount of time with high frequency measurements. To see how the orbit determination errors grow with low frequency measurements, Monte Carlo simulations with 100 executions have been performed for various frequencies. This has been done based on the link between the Earth-Moon L2, L1 Halo Orbiters and Lunar Orbiter with \SI{1}{m} ranging accuracy. Measurement intervals have been set as \SI{5e-4}{TU} (\SI{187}{s}), \SI{2.5e-3}{TU} (\SI{938}{s}), and \SI{5e-3}{TU} (\SI{1876}{s}). Corresponding results can be seen in Figure~\ref{fig:measfreq}.

As might be expected, high measurement intervals increase the \ac{RMS} measurement error. On the other hand, the EM L1-Lunar case showed that less measurements do not always mean having higher estimation errors. Basically, in the \SI{1870}{s} measurement interval case, measurements are collected in such a relative geometry that the navigation filter has been fed by more useful data than the case with a \SI{938}{s} measurement interval case. In addition, orbits with shorter periods (the Lunar Orbiter in these cases) are less sensitive to measurement interval changes. 

\subsection{Formation Geometry}

In this section, the effect of the relative geometry between satellites has been investigated. This was done by changing the orbital period of the Lagrangian orbiters at the Earth-Moon L1 and L2. This eventually affects the inter-satellite distance and thus the relative geometry. Different constellation geometries, coplanar and non-coplanar cases at the Lagrangian points,  have been investigated in \cite{hillthesis} and it was shown that the autonomous orbit determination would not work for close formations but only for those with large separations. This section will study the formation geometry from the observability perspective for formations located at different Lagrangian points.

\begin{figure}[h]
    \centering
    \includegraphics[width=0.33\textheight]{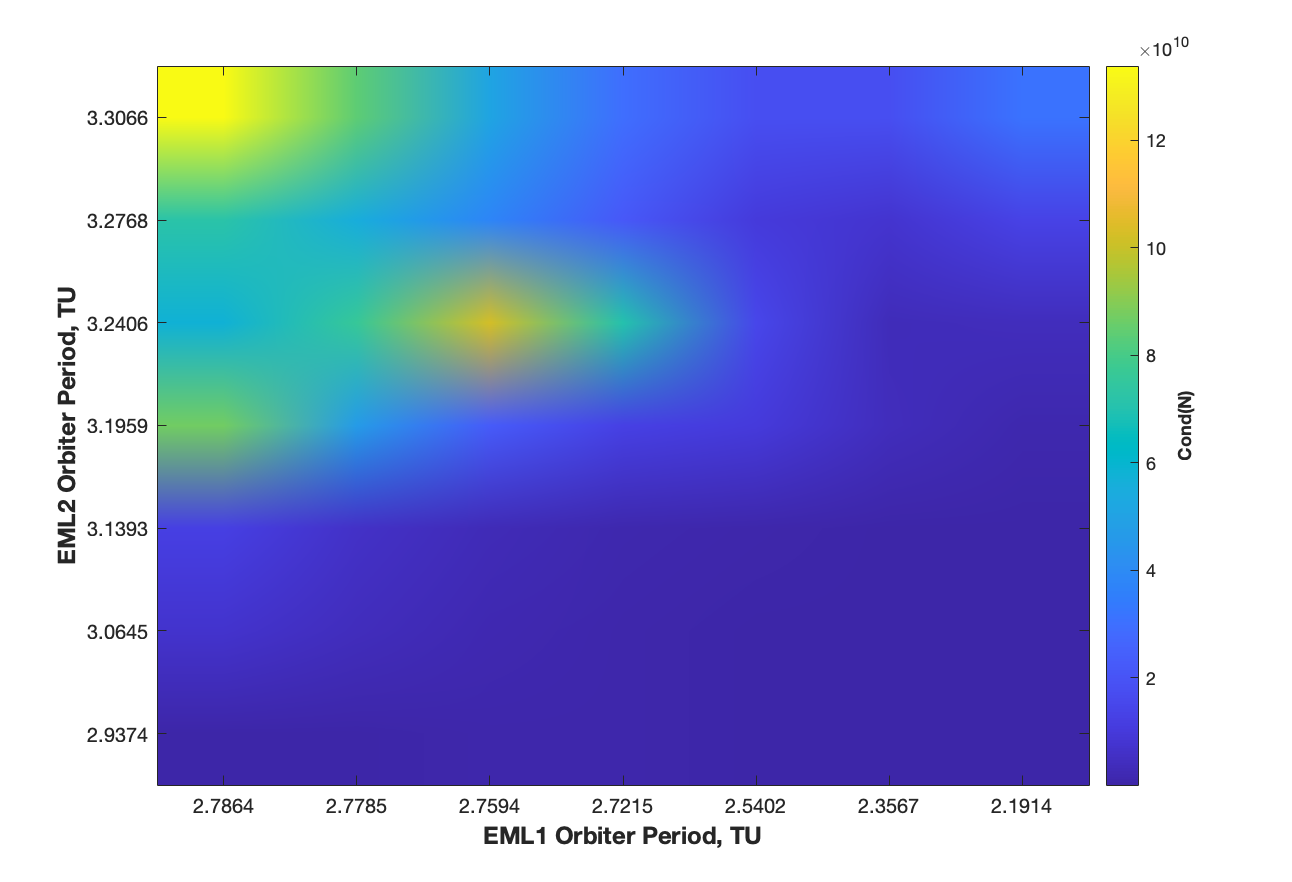}
    \caption{Condition number variation among various L1 and L2 Orbiters}
    \label{fig:condfg}
\end{figure}

\begin{figure}[h]
    \centering
    \includegraphics[width=0.33\textheight]{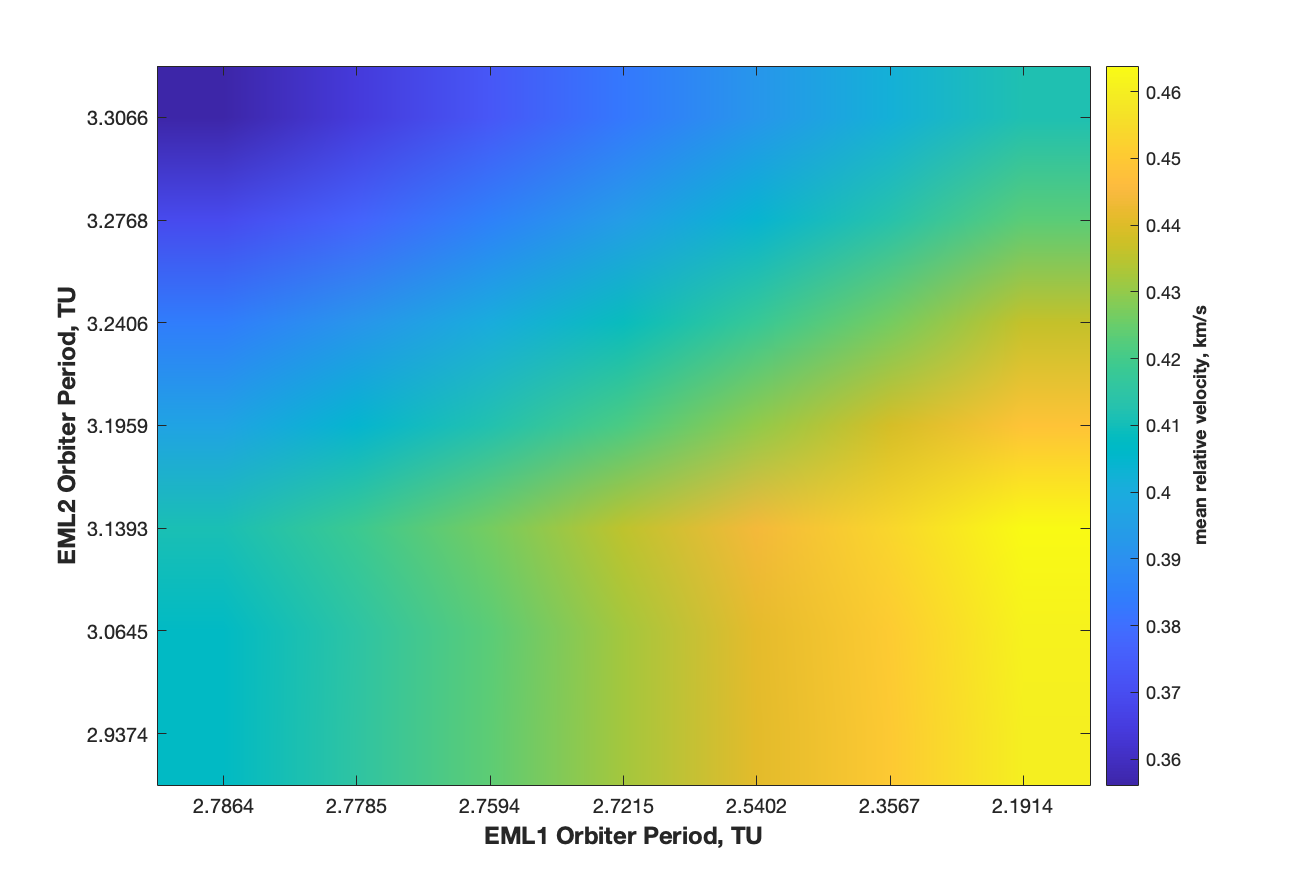}
    \caption{Mean relative velocity between spacecraft among various L1 and L2 Orbiters}
    \label{fig:condfg2}
\end{figure}

\begin{figure}[h]
    \centering
    \includegraphics[width=0.33\textheight]{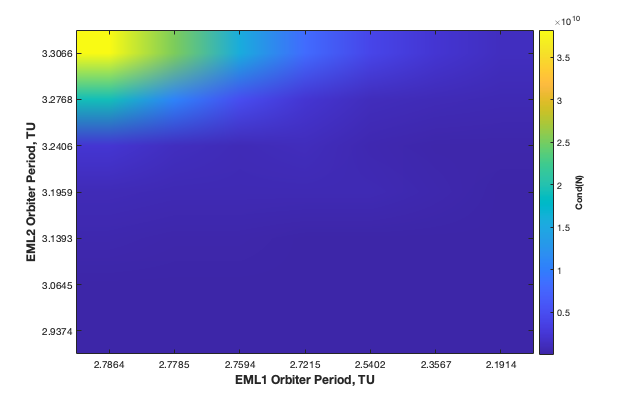}
    \caption{Condition number variation among various L1 and L2 Orbiters based on range-rate measurements}
    \label{fig:condfg3}
\end{figure}

Simulations have been configured such that one of the spacecraft has an EM L1 Halo orbit and the second one has an EM L2 Halo orbit. In the first case, all possible crosslink scenarios between seven EM L1 and EM L2 Halo orbiters have been simulated. In these cases, the inter-satellite distances have been changed from a minimum of \SI{18411}{km} to a maximum of \SI{133738}{km}. The duration of the simulations has been adjusted based on the period of the spacecraft with the longest orbital period, ranging between $12.75$ days and $14.36$ days. Figure~\ref{fig:condfg} shows the condition number variation among the various L1 and L2 orbiters. It can be seen that the system becomes less observable when both spacecraft have longer orbital periods and more observable when both have shorter periods or one of them has a short and other one has a longer period. Basically, shorter orbital periods have higher relative velocity (see the Figure~\ref{fig:condfg2} for further details) and spacecraft with shorter period can obtain more information while spacecraft with a longer period completes its one orbital period. On the other hand, Lagrangian orbiters with shorter periods are in general close to the Moon and their orbital plane becomes planar, allowing to collect more information in the x-y plane. In the case of range-rate observations, the results did not change from those provided in Figure~\ref{fig:condfg3}.

\begin{figure}[h]
    \centering
    \includegraphics[width=0.33\textheight]{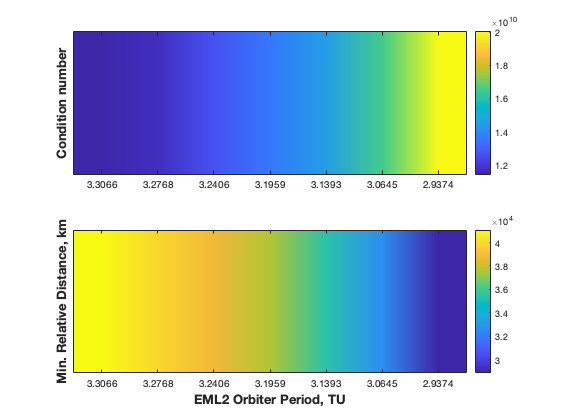}
    \caption{Condition number and minimum relative distance variation among various L2 and Lunar Orbiter}
    \label{fig:condfg4}
\end{figure}

In the second run, the crosslink between EM L2 Halo orbiter and the Lunar orbiter has been investigated. The same trends have been observed in this case and the corresponding results are provided in Figure~\ref{fig:condfg4}. 

\subsection{Network Topology}

In case there are more than two spacecraft in the system, the network topology would become an important parameter to consider. Increasing the number of spacecraft, thus increasing the number of inter-satellite links, would improve the overall navigation performances. This is due to the increase in information acquired at the same time, thanks to the different relative geometries between spacecraft. In this section, triple spacecraft scenarios have been investigated considering two different network topologies: centralized (star) and distributed (mesh). The centralized, or in other words star topology, has the advantage of simplicity but has a single-point-of-failure. On the other hand, distributed or mesh topologies supports interactions among all spacecraft but they suffer from complexity as the number of spacecraft and links increases. These different scenarios are listed in Table~\ref{triplecentralized} and~\ref{triplemesh}.

\begin{table}[h]
\centering
\caption{Triple Spacecraft Scenarios for the Centralized Topology}
\label{triplecentralized}
\begin{tabular}{|c|c|c|} 
\hline
\multirow{2}{*}{Scenario ID} & \multicolumn{2}{c|}{Triple Spacecraft Configurations -~Centralized Topology~} \\ 
\cline{2-3}
 & Mothercraft & Daughtercrafts \\ 
\hline
C1 & EML2 Halo & EML1 Halo - Lunar Orbiter \\ 
\hline
C2 & EML1 Halo & EML2 Halo - Lunar Orbiter \\ 
\hline
C3 & Lunar Orbiter & EML2 Halo - EML1 Halo Orbiter \\ 
\hline
C4 & EML2 Halo & EML2 Halo (non-coplanar) - Lunar Orbiter \\ 
\hline
C5 & EML2 Halo & EML2 NRHO - Lunar Orbiter \\ 
\hline
C6 & EML2 Halo & EML2 Lyapunov - Lunar Orbiter \\
\hline
\end{tabular}
\end{table}

\begin{table}[h]
\centering
\caption{Triple Spacecraft Configurations - Mesh Topology}
\label{triplemesh}
\begin{tabular}{|c|l|} 
\hline
Scenario ID & Triple Spacecraft Configurations - Mesh Topology \\ 
\hline
M1 & EML2 Halo - EML1 Halo - Lunar Orbiter \\ 
\hline
M2 & EML2 Halo - EML2 Halo (non-coplanar) - Lunar Orbiter \\ 
\hline
M3 & EML2 Halo - EML2 NRHO - Lunar Orbiter \\ 
\hline
M4 & EML2 Halo - EML2 Lyapunov - Lunar Orbiter \\
\hline
\end{tabular}
\end{table}

In the centralized topology, the very first three scenarios have the same relative geometry but different centralized node (mothercraft), which results in different inter-satellite measurement vectors. In other words, the mothercraft collects observations in each centralized scenario. This will show the effect of the orbital period on the overall navigation performances. The last three scenarios have the same spacecraft as a central node but different configurations of daughtercraft. This will allow to see the effect of different relative geometries/inter-satellite distances. 

In the mesh topology scenarios, all the spacecraft have measurements between each others. The first scenario is nothing but the mesh version of the very first three scenarios previously presented. The last three scenarios are the mesh version of the corresponding centralized topology. These scenarios will show the effect of having additional inter-satellite link on the overall navigation performance. 

\begin{figure}[h]
    \centering
    \includegraphics[width=0.33\textheight]{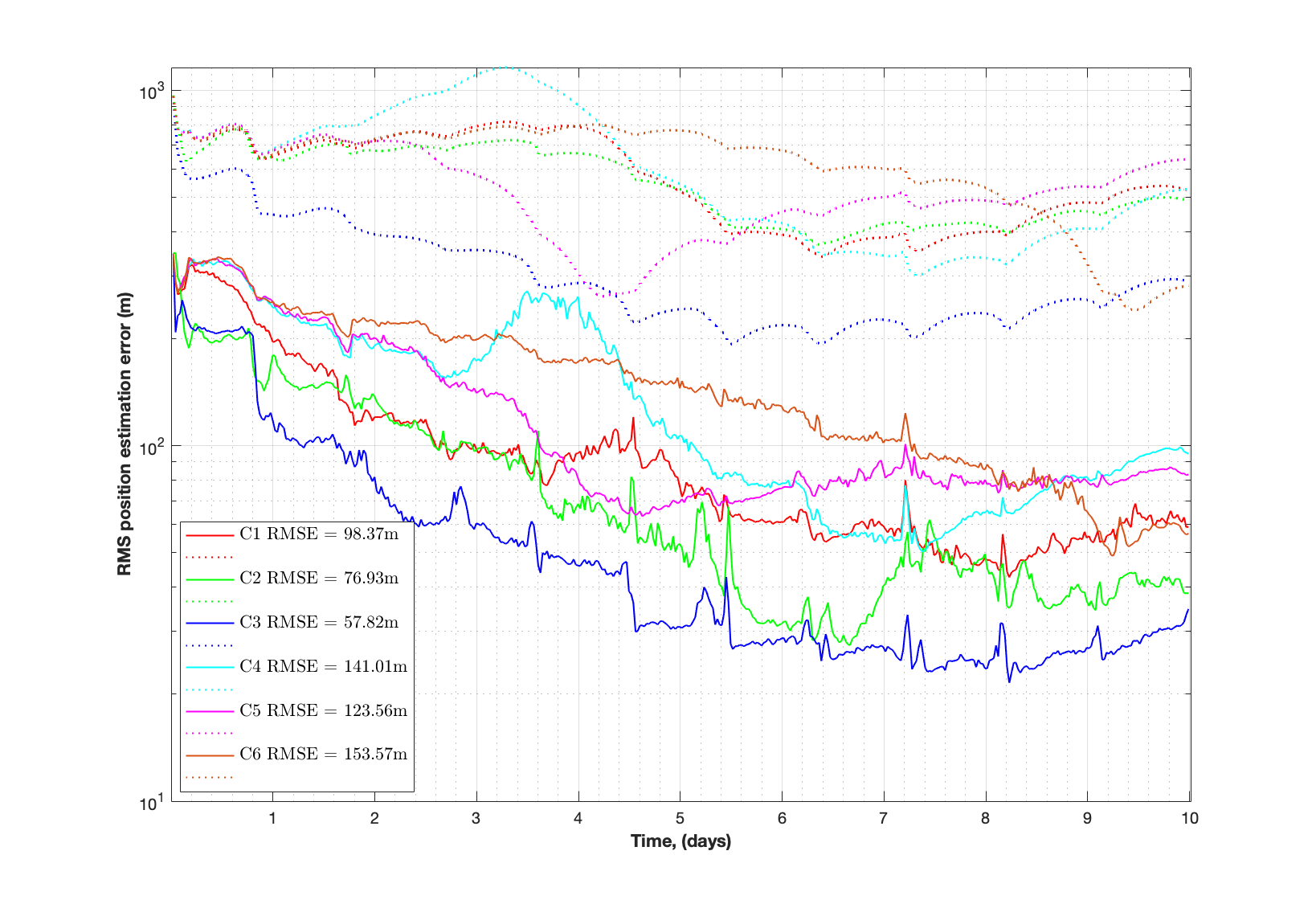}
    \caption{RMS position estimation error for centralized topology scenarios}
    \label{fig:Cpos}
\end{figure}

\begin{figure}[h]
    \centering
    \includegraphics[width=0.33\textheight]{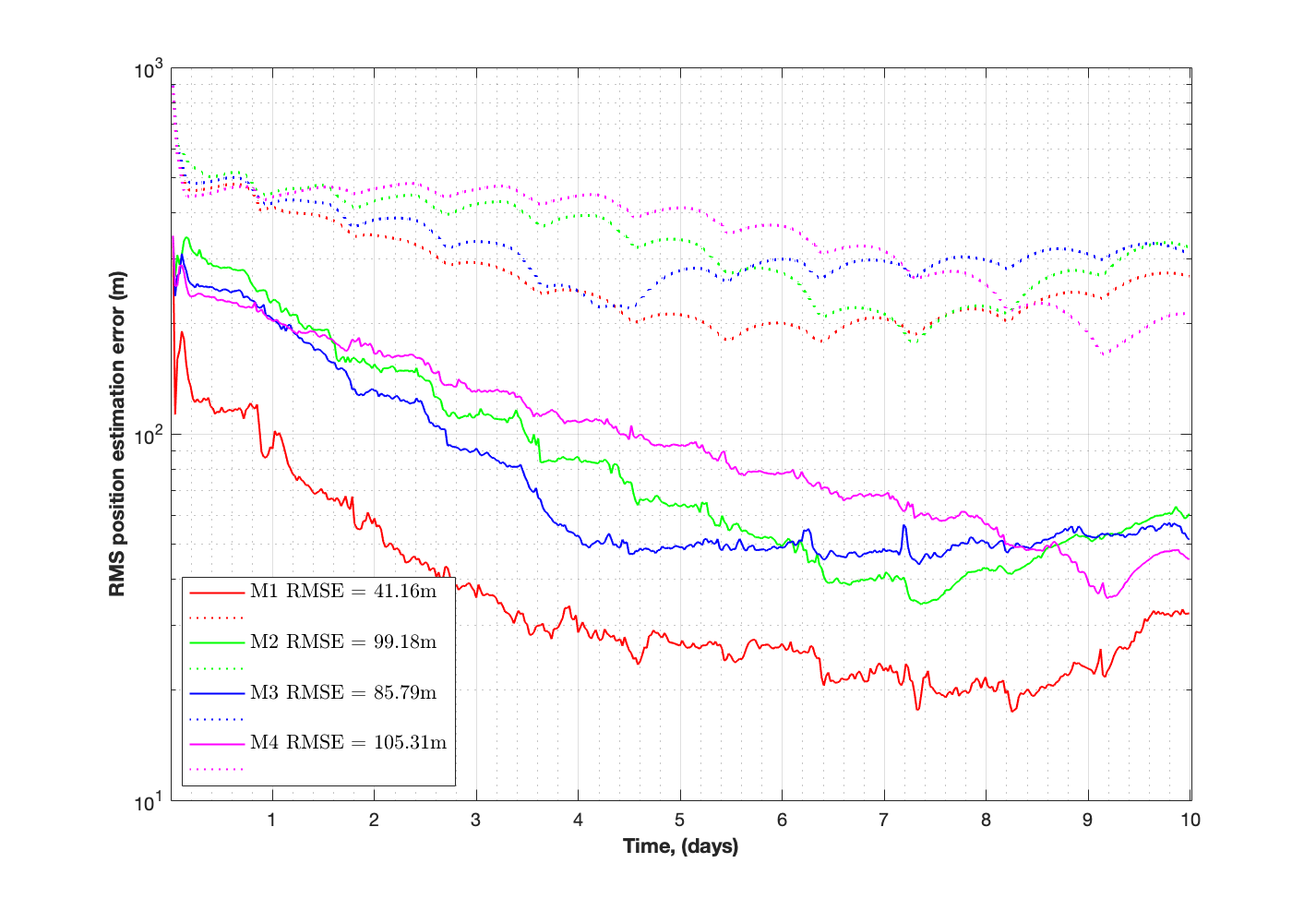}
    \caption{RMS position estimation error for mesh topology scenarios}
    \label{fig:Mpos}
\end{figure}

All the centralized and mesh topology \ac{RMS} position estimation errors (three S/C position errors have been averaged over 100 Monte Carlo executions) can be seen in Figure~\ref{fig:Cpos} and~\ref{fig:Mpos} respectively. \ac{RMS} position and velocity estimation errors for both centralized and mesh topologies can be seen in Table~\ref{tab:Centralized} and Table~\ref{tab:Mesh}. The unobservability index for the centralized and mesh topology scenarios can be seen in Table~\ref{unobsC} and~\ref{unobsM}, respectively. 

\begin{table}
\centering
\caption{RMS Estimation errors for various centralized topologies}
\begin{tabular}{|l|c|c|c|c|c|c|} 
\hline
 & C1 & C2 & C3 & C4 & C5 & C6 \\ 
\hline
RMS Pos.Error (m) & 98.37 & 76.93 & 57.82 & 141.01 & 123.56 & 153.57 \\ 
\hline
RMS Vel.Error (mm/s) & 1.15 & 0.97 & 0.71 & 1.43 & 1.32 & 1.37 \\
\hline
\end{tabular}
\label{tab:Centralized}
\end{table}

\begin{table}
\centering
\caption{RMS Estimation errors for various mesh topologies}
\begin{tabular}{|l|c|c|c|c|} 
\hline
 & M1 & M2 & M3 & M4 \\ 
\hline
RMS Pos.Error (m) & 41.16 & 99.18 & 85.79 & 105.31 \\ 
\hline
RMS Vel.Error (mm/s) & 0.63 & 1.03 & 0.88 & 0.91 \\
\hline
\end{tabular}
\label{tab:Mesh}
\end{table}

Considering the very first three centralized scenarios (C1,C2,C3), a centralized node with a shorter orbital period, lunar orbiter (C3), provides better estimation accuracy than a mothercraft located at the Lagrangian points (C1, C2). This is due to the fact that Lunar orbiter had better observation geometry and collected all the information in a shorter time. In addition, this case (C3) has a lower unobservability index than the other cases. On the other hand, the results of the last three centralized scenarios showed that larger inter-satellite distances provide better state estimation. Comparing C2 with C4 shows that having a shorter inter-satellite link in the topology decreases the overall navigation system performances.

Regarding the mesh scenarios, the first case (M1) showed that adding an additional link, thus measurement, on a centralized scenario (C1,C2,C3) improved the navigation performances. This behaviour has been observed in other cases as well by comparing M2, M3, and M4 with C2, C3 and C4. As it can be seen, M3 converges faster than M4 showing that shorter orbital periods (T=\SI{7.94}{days}, T=\SI{18.71}{days}) in the formation provide a quicker solution. The reason is that the filter in this case is fed by observations containing information about the full orbital trajectory and not just a sub-section.

\begin{table}[h]
\centering
\caption{Unobservability index for centralized topology scenarios}
\label{unobsC}
\begin{tabular}{|c|c|c|c|c|c|c|} 
\hline
 & C1 & C2 & C3 & C4 & C5 & C6  \\ 
\hline
$1/\text{min}(eig(\boldsymbol{N}))$ & \SI{2.97e2}{}  & \SI{7.74e2}{}  & \SI{1.91e2}{}  & \SI{4.5e2}{} & \SI{3.54e2}{} & \SI{3.85e2}{}  \\
\hline
\end{tabular}
\end{table}

\begin{table}[h]
\centering
\caption{Unobservability index for mesh topology scenarios}
\label{unobsM}
\begin{tabular}{|c|c|c|c|c|} 
\hline
& M1 & M2 & M3 & M4 \\ 
\hline
$1/\text{min}(eig(\boldsymbol{N}))$ & \SI{1.25e2}{} & \SI{1.52e2}{}  & \SI{1.36e2}{}  & \SI{1.96e2}{}  \\
\hline
\end{tabular}
\end{table}

\section{Conclusion}

The aim of this study was to investigate the effect of different aspects on the performances of a crosslink radiometric measurement based autonomous navigation method, \ac{LiAISON}, for the special case of cislunar small satellite formations. Various crosslink radiometric data types have been presented including their expected performances. Thereafter,  orbit determination models and various performance analysis methods were also given. The navigation system performances have been studied to quantify the effects of important parameters including measurement type, accuracy, bias, frequency, formation geometry, and network topology. It was found that range observations provide a better state estimation than range-rate observables for the autonomous navigation system in cislunar space. Adding \ac{LOS} measurements into the filter provided better results for selected mission scenarios only if \ac{LOS} measurements are accurate (better than \SI{3.6}{arcsec}), which is difficult to achieve in practice via radio-interferometric methods on small satellites. On the other hand, combining range and range-rate observables slightly improved the performances. Instead of combining them, one approach would be investigating the observation effectiveness for each observation type, and using range or range-rate only measurements in the best tracking windows. About the measurement accuracy, high-observable systems have been coped with high measurement errors. This means that less accurate inter-satellite measurement methods, such as time-derived or data-aided ranging, could be an option for these formations. As might be expected, high measurement intervals increase the \ac{RMS} measurement error. About the formation geometry, orbiters with two shorter periods or one shorter and one longer periods presented better observable systems. In case there are more than two spacecraft in the system, the mesh topology provided more accurate state estimations than a centralized topology, as expected, because of the increased number of measurements. However, considering the overall system performances, it would be beneficial for the centralized topology to collect all the measurements on the spacecraft with the shortest orbital period. 

\appendix
\section{}
\label{appendixa}

In this section, the variance for line-of-sight measurements is given based on both a time-delay and phase-shift approach. 

The variance of a continuous random variable X is defined as:
\begin{eqnarray}
    \text{Var}[X]=\text{E}[(X-\text{E}[X])^2]=\text{E}[X^2]-(\text{E}[X])^2
\end{eqnarray}
for constants $a$ and $b$:
\begin{eqnarray}
    \text{Var}[a+bX]=b^2\text{Var}[X]
\end{eqnarray}
Taking the variance of both hand sides of Eq.~\ref{eqndeltat} for the time-delay based approach:
\begin{eqnarray}
    \text{Var}[\Delta t]=\frac{b^2}{c^2}\text{Var}[\cos \psi]
\end{eqnarray}

Variance of $\cos \psi$ is given as:
\begin{eqnarray}
\label{varcospsi}
    \text{Var}[\cos \psi]=\text{E}[\cos^2 \psi]-(\text{E}[\cos \psi])^2
\end{eqnarray}

where
\begin{eqnarray}
\label{cospsi}
    \text{E}[\cos \psi]= \sum_{k=0}^{\infty} \frac{(-1)^k}{2k!} \text{E}[\psi^{2k}] = \sum_{k=0}^{\infty} \frac{(-1)^k}{2k!} \sigma ^{2k} (2k-1)! = e^{-\sigma_{\psi}^{2}}
\end{eqnarray}

and
\begin{eqnarray}
    \text{E}[\cos^2 \psi] = 1 - \text{E}[\sin^2 \psi] = 1 - (\frac{1}{2}(1-\text{E}[\cos 2\psi])) = 1 - (\frac{1}{2}(1-\sum_{k=0}^{\infty} (-1)^k \frac{2^{2k}}{2k!} \text{E}[\psi^{2k}]))
\end{eqnarray}

\begin{eqnarray}
\label{cossqrpsi}
    \text{E}[cos^2 \psi] = 1 - (\frac{1}{2}(1-\sum_{k=0}^{\infty} (-1)^k \frac{2^{2k}}{2k!} \sigma ^{2k} (2k-1)!)) = 1 - (\frac{1-e^{-2\sigma_{\psi}^{2}}}{2}) = (\frac{1+e^{-2\sigma_{\psi}^{2}}}{2})
\end{eqnarray}

Inserting Eq.~\ref{cospsi} and~\ref{cossqrpsi} into~\ref{varcospsi} results in:
\begin{eqnarray}
    \text{Var}[\Delta t]= \frac{b^2}{c^2} ((\frac{1+e^{-2\sigma_{\psi}^{2}}}{2}) - e^{-2\sigma_{\psi}^{2}}) = \frac{b^2}{c^2} (\frac{1-e^{-2\sigma_{\psi}^{2}}}{2})
\end{eqnarray}

This can also be written in terms of ranging error:
\begin{eqnarray}
    \sigma_{\rho}= b \sqrt{\frac{1-e^{-2\sigma_{\psi}^{2}}}{2}}
\end{eqnarray}

or
\begin{eqnarray}
    \sigma_{\psi}= \sqrt{ln \left (\frac{1}{\sqrt{1-\frac{2\sigma_{\rho}^{2}} {b^2}}} \right )}
\end{eqnarray}

Similarly, the relation between phase-shift measurement error and line-of-sight measurement error can be found by taking the variance of both hand sides of Eq. (\ref{eqntau}):
\begin{eqnarray}
    \text{Var}[\tau]=\frac{(2\pi b)^2}{\lambda^2}\text{Var}[\sin \psi]
\end{eqnarray}

Since a zero-mean Gaussian with variance $\sigma_{\psi}^2$ is assumed ($\text{E}[\sin \psi]=0$), the variance of $\sin \psi$ is given as:
\begin{eqnarray}
\label{varcospsi1}
    \text{Var}[\sin \psi]=\text{E}[\sin^2 \psi] = \frac{1}{2} (1- \text{E}[\cos (2\psi)])=\frac{1-e^{-2\sigma_{\psi}^{2}}}{2}
\end{eqnarray}

This can be written as:
\begin{eqnarray}
    \sigma_\tau = \frac{2\pi b}{\lambda} \sqrt{\frac{1-e^{-2\sigma_{\psi}^{2}}}{2}}
\end{eqnarray}
or
\begin{eqnarray}
    \sigma_{\psi}= \sqrt{ln \left (\frac{1}{\sqrt{1-\frac{2\sigma_{\rho}^{2} \lambda^2} {(2 \pi b)^2}}} \right )}
\end{eqnarray}

It is also can be seen in the phase-shift measurement approach that line-of-sight measurements would not be possible if $\sigma_\tau < \frac{\sqrt{2}\pi b}{\lambda}$ is not met. As a side note, time-delay and phase-shift measurements can be invertable as $\sigma_{\Delta t}=\sigma_{\tau} \lambda / 2 \pi c$

\section{}
\label{appendixb}
In this section, the state transition matrix generated by the numerical integrator (ODE113 in Matlab) used as a reference is compared with the state transition matrix approximated by the method given in subsection~\ref{subsecobservability}. Based on the following relative error parameter, 
\begin{eqnarray}
    \epsilon =\frac{1}{36}\sum_{i=1}^{6}\sum_{j=1}^{6}\left | \frac{(\mathbf{\Phi}_{ij} - \mathbf{\bar\Phi}_{ij})}{\boldsymbol{\bar\Phi}_{ij}} \right |
\end{eqnarray}

where $\mathbf{\bar\Phi}_{ij}$ and $\mathbf{\Phi}_{ij}$ represent the \ac{STM} calculated by the numerical integrator and approximated method respectively. A comparison has been made based on two different orbits, namely Earth-Moon L2 and Earth-Moon L1 Southern Halo for 7 days of duration. \ac{RMS} errors and their standard deviations for various step sizes are given in Table~\ref{stmtable}.
\begin{table}[h]
\centering
\caption{Performances of the approximated \ac{STM}.}
\label{stmtable}
\begin{tabular}{|c|c|c|c|c|} 
\hline
\multirow{2}{*}{$\Delta t (s)$} & \multicolumn{2}{c|}{$\epsilon$ for EM L2 orbiter} & \multicolumn{2}{c|}{$\epsilon$ for EM L1 orbiter} \\ 
\cline{2-5}
 & RMSE & $1\sigma$ STD & RMSE & $1\sigma$ STD \\ 
\hline
10 & \SI{9.29e-10}{} & \SI{7.54e-10}{} & \SI{1.75e-9}{} & \SI{1.13e-9}{} \\ 
\hline
60 & \SI{6.44e-9}{} & \SI{1.64e-8}{} & \SI{1.22e-8}{} & \SI{2.83e-8}{} \\ 
\hline
100 & \SI{1.42e-8}{} & \SI{3.89e-8}{} & \SI{2.65e-8}{} & \SI{6.70e-8}{} \\ 
\hline
600 & \SI{6.95e-8}{} & \SI{5.54e-7}{} & \SI{1.31e-7}{} & \SI{9.68e-7}{} \\
\hline
\end{tabular}
\end{table}


\bibliographystyle{model5-names}
\biboptions{authoryear}
\bibliography{article2}

\end{document}